\documentclass{elsart}

\usepackage{amsmath}
\usepackage{graphicx}
\usepackage{multirow}
\usepackage{cite}

\journal{Surface Science}

\begin{document}

\begin{frontmatter}

\title{Kinetic Monte Carlo simulations of electrodeposition: Crossover
from continuous to instantaneous homogeneous nucleation within Avrami's law}

\author[Phys,Martec,SCS]{Stefan Frank\thanksref{now}}
\thanks[now]{Present address: Zentrum f\"ur Sonnenenergie- und
  Wasserstoff-Forschung Baden-W\"urttemberg, Helmholtzstra{\ss}e 8,
  89081 Ulm, Germany, and Universit\"at Ulm, Abteilung Theoretische
  Chemie, 89069 Ulm, Germany.}
\ead{stefan.frank@uni-ulm.de}
\author[Phys,Martec,SCS,Maglab]{Per Arne Rikvold}
\ead{rikvold@scs.fsu.edu}
\ead[url]{http://www.physics.fsu.edu/users/rikvold/info/rikvold.htm}

\address[Phys]{Department of Physics, Florida State University,
  Tallahassee, FL 32306-4350, USA}
\address[Martec]{Center for Materials Research
  and Technology, Florida State University, Tallahassee, FL 32306-4351, USA}
\address[SCS]{School of Computational Science, Florida State
  University, Tallahassee FL 32306-4120, USA}
\address[Maglab]{National High Magnetic Field Laboratory, Tallahassee, FL 32310-3706, USA}

\begin{abstract}
The influence of lateral adsorbate diffusion on the dynamics of the
first-order phase transition in a two-dimensional Ising lattice gas
with attractive nearest-neighbor interactions is investigated by means
of kinetic Monte Carlo simulations. For example, electrochemical
underpotential deposition proceeds by this mechanism. One major
difference from adsorption in vacuum surface science is that under
control of the electrode potential and in the absence of
mass-transport limitations, local adsorption equilibrium is
approximately established. We analyze our results using the theory of
Kolmogorov, Johnson and Mehl, and Avrami (KJMA), which we extend to an
exponentially decaying nucleation rate. Such a decay may occur due to
a suppression of nucleation around existing clusters in the presence
of lateral adsorbate diffusion. Correlation functions prove the
existence of such exclusion zones. By comparison with microscopic
results for the nucleation rate~$I$ and the interface velocity of the
growing clusters~$v$, we can show that the KJMA theory yields the
correct order of magnitude for $Iv^2$. This is true even though the
spatial correlations mediated by diffusion are neglected. The decaying
nucleation rate causes a gradual crossover from continuous to
instantaneous nucleation, which is complete when the decay of the
nucleation rate is very
fast on the time scale of the phase transformation. Hence,
instantaneous nucleation can be homogeneous, producing negative minima
in the two-point correlation functions. We also present in this paper
an $n$-fold way Monte Carlo algorithm for a square lattice gas with
adsorption/desorption \textit{and} lateral diffusion.
\end{abstract}


\end{frontmatter}

\section{Introduction}
\label{par:introduction}

The electrochemical adsorption of a two-dimensional film on a metal
electrode can involve a phase transition at the
surface~\cite{Blum96}. When the adsorbate--adsorbate interactions are
attractive, this phase transition is of first order; hence, below a
critical temperature, there is a discontinuity in the coverage at a
certain electrochemical potential. Stepping across the electrostatic
potential difference between electrode and solution that corresponds
to this coexistence point induces adsorption (or desorption). Whereas,
thermodynamically, this phase transformation is abrupt rather than
gradual, its kinetics are governed by the microscopic mechanism by
which it proceeds. On a flat, defect-free single-crystal surface, the
decay of the metastable phase takes place through nucleation and
growth of two-dimensional clusters. The classical theory for this kind
of phase change was developed in the 1930s and '40s by Kolmogorov,
Johnson and Mehl, and Avrami (KJMA
theory)~\cite{Kolmogorov37,Johnson39,Avrami39,Avrami40,Avrami41}.
Finding successful application in many fields of science, it is also
the basis for the interpretation of kinetic data of electrodeposition
processes that proceed by nucleation and
growth~\cite{SchmicklerBook96,Holzle94}. However, the assumptions of
the theory may not always be met in real experimental systems. In
particular, an earlier study from our group indicates that homogeneous
nucleation can also be instantaneous, and that there might be a
gradual crossover from continuous to instantaneous
nucleation~\cite{Frank05}. This casts some doubt on whether the
nucleation is spatially uncorrelated~\cite{Pineda02,Fanfoni03}.

Lattice-gas models are now a widely applied tool in the study of
electrochemical adsorption~\cite{BrownBook99,RikvoldBook02}. They are
appropriate when the adsorbate layer is commensurate with the
substrate~\cite{Rojas04}. When the adsorbate is in quasi-equilibrium
with the bulk under control of the electrode potential, a
grand-canonical Hamiltonian should be used. Kinetic Monte Carlo
simulations~\cite{BinderBook97} of lattice-gas models can be used to
study the dynamics of such systems when they can be well approximated
as a superposition of independent, stochastic
processes~\cite{BrownBook99}. This is generally true when the
elementary processes evolve through thermally activated crossings of
well-defined transition states. Then, one should choose an
Arrhenius-type Monte Carlo dynamic~\cite{Kang89,Fichthorn91}. For the
investigation of the model dynamics at macroscopic time scales this is
the method of choice, since it does not rely on mathematical
approximations and since it coarse-grains over the fluctuations (vibrations)
whose time scales are orders of magnitude faster than adsorption,
diffusion, etc. With this method real problems have been tackled and
successfully described~\cite{Abou04}, for example for underpotential
deposition of metals~\cite{AramataBook97, Brown99a}. However, the problem may
require substantial effort in providing a sufficiently realistic
description. For example, one may need to include coadsorption of
anions, long-range or many-body potentials~\cite{Abou04,Rojas04,Luque05}, etc.

Here we follow a somewhat different approach. Rather than in a
specific system, we are interested in the general behavior of
two-dimensional first-order phase transitions. Hence we choose as a
model system a two-dimensional square lattice gas with
nearest-neighbor attractive interactions. No solvent or counter ions
are included. This model can be interpreted as a simplified
description of electrochemical metal underpotential deposition or of
the early stages of epitaxial growth when no nucleation takes place in
the second layer. However, since it is isomorphic to the
square-lattice Ising model~\cite{RikvoldBook94}, it is far more
general and may, among other phenomena, also describe ferromagnetic
systems~\cite{Ramos99}. The focus of this paper is on the influence of
nearest-neighbor adsorbate diffusion on the dynamics of the phase
transformation. Unlike in vacuum surface science, adsorbate diffusion
is not a condition for nucleation since fluctuations in cluster sizes
can also be mediated by the quasi-equilibrium with the solution
(instead of the two-dimensional gas of adsorbed particles). Likewise,
an earlier study~\cite{Ramos99} has shown that the
phase-transformation dynamics of this lattice gas can be well
described by the KJMA theory for continuous nucleation over a wide
range of potentials in the \textit{absence} of adsorbate diffusion
(in magnetic terms, spin flip only). Another preliminary
study~\cite{Frank05} suggests that in the \textit{presence} of
diffusion (spin exchange in magnetic language) the dynamics may
change. For fast diffusion, the phase transformation appeared to
proceed by instantaneous nucleation, and for intermediate diffusion
rates there was a gradual crossover. This was interpreted as a result
of exclusion zones around the clusters, in which nucleation was
suppressed. It was speculated that these exclusion zones may become
space-filling in the early stages of the phase transformation. The
crossover also changed the morphology of the adsorbate phase. However,
the study used Glauber instead of Arrhenius dynamics and was limited
to one value of the electrochemical potential. Here we extend the
study to a wide range of potentials. Furthermore, we present a more
stringent analysis of the kinetic data and use microscopic details of
the simulation for comparison with the theory.

The remainder of this paper is organized as follows. In
Sec.~\ref{par:theo} we review the KJMA theory and extend it to an
exponentially decaying nucleation rate. In Sec.~\ref{par:model} we
introduce the lattice-gas model and the details of the algorithm and
the model calculations. Results are presented thereafter. In
Sec.~\ref{par:results:dynamics} we describe the phase transformation
dynamics and the morphology of the adsorbate phase and give kinetic
parameters according to the theory. In Sec.~\ref{par:results:micro} we
present the results for the microscopic rates. In
Sec.~\ref{par:results:continuum} we discuss the accuracy of the
continuum description of the KJMA theory, and in
Sec.~\ref{par:results:onsetcrover} we discuss under which conditions
the crossover to instantaneous nucleation occurs. A summary and
conclusions are given in Sec.~\ref{par:summary}. A description of the
$n$-fold way algorithm for adsorption/desorption \textit{and}
diffusion can be found in the Appendix.

\section{Theory}
\label{par:theo}

We here briefly review the KJMA theory and describe how we
adapt it for a gradual crossover between continuous and instantaneous
nucleation. 

The basic assumption of the KJMA theory is that nucleation events are
\textit{random} and \textit{uncorrelated}. Then, it is mathematically
quite simple to correct for the overlap of growing circular droplets,
provided that they do not change shape upon
coalescence~\cite{Evans45,Fanfoni96}. 
One defines an \textit{extended
coverage}~$\theta_{\mathrm{ext}}$ by summing up the areas of all
growing droplets regardless of overlap with other clusters, and
normalizes this by the total surface area. When the radius~$r$ of a
cluster is a function of its birth time~$\tilde{t}$, this reads
\begin{equation}
\theta_{\mathrm{ext}} = \int_0^{r_\mathrm{max}} \rho(r) \pi r^2 \mathrm{d} r = \pi v^2
\int_0^t{I(\tilde{t})(t-\tilde{t})^2}\mathrm{d}\tilde{t},\label{eq:ext_cov}
\end{equation}
where $\rho(r)$ is the number of clusters of radius~$r$ per surface
area, and $I(t)$ is the nucleation rate at time~$t$. The
last equality is for circular clusters with a radius growing linearly
with an interface velocity~$v$. It is important to realize that this
nucleation rate includes nucleation events in the transformed area,
since otherwise the condition of random, uncorrelated nucleation would
be violated~\cite{Sessa96,Tomellini97}. There has been some confusion
concerning this point~\cite{Erukhimovitch95}.  We therefore call the
nucleation rate~$I(t)$ the \textit{extended nucleation rate}
wherever the distinction is important. In the system we consider in
our investigation, nucleation is homogeneous and proceeds by thermal
fluctuations. It can be described by classical nucleation
theory~\cite{RikvoldBook94}, but we do not address this question further here. 

When $I$ is constant throughout the phase transformation, we speak of
\textit{continuous nucleation}, and performing the integration in
Eq.~(\ref{eq:ext_cov}) yields
\begin{equation}
\theta_{\mathrm{ext}} = \frac{\pi I v^2 t^3}{3}.\label{eq:ext_cov_cont}
\end{equation}
Here, $I$ has the dimension $1/(\mathrm{area}\times \mathrm{time})$.
It is clear that the ``real'' nucleation rate~$I_{\mathrm{r}}$ is
reduced by the fraction of the covered surface $\theta$ and eventually
decreases to zero; this is not to be confused with a decaying extended
nucleation rate (see below). At early times, though, the real
nucleation rate can be used as an estimate for the extended nucleation
rate, since $\theta \approx 0$. 

In contrast, when all nuclei are already present at the very beginning
of the phase transformation and no further nucleation takes place,
$I(t)=I \delta(t)$, where $\delta(t)$
is a Dirac delta function, and the integration in Eq.~(\ref{eq:ext_cov})
yields
\begin{equation}
\theta_{\mathrm{ext}} = \pi I v^2 t^2.\label{eq:ext_cov_inst}
\end{equation}
This \textit{instantaneous nucleation} is often associated with
heterogeneous nucleation at defects. Here, $I$ has the dimension
$1/\mathrm{area}$ since it is virtually a surface density of
nucleation centers. 

The real coverage can be calculated from the extended coverage~$\theta$
by~\cite{Avrami40}
\begin{equation}
1-\theta= \phi_{\mathrm{KJMA}} =\mathrm{e}^{-\theta_{\mathrm{ext}}}.\label{eq:avrami}
\end{equation}
The quantity $\phi_{\mathrm{KJMA}}$ is called the relaxation function
(in the KJMA theory). We will later define a relaxation function
for our model system and use $\phi_{\mathrm{KJMA}}$ as a
coarse-grained approximation for it~\cite{Ramos99}.

\subsection{Extended volume for decaying nucleation rate}
\label{par:theo:decaynucl}
The gradual crossover from continuous to instantaneous nucleation with
increasing diffusion rate implies that, for some intermediate
diffusion rates, the extended nucleation rate~$I$ is neither a
constant nor a delta function, but decays with time. In the following,
we deduce how an exponentially decaying extended nucleation rate
\begin{equation}
I(t) = I_0 \mathrm{e}^{-\lambda t} \label{eq:decay_nucl}
\end{equation}
with $I_0$ the initial nucleation rate and $\lambda$ an inverse decay
time translates into a $\theta(t)$ relation for the coverage. The
dimension of $I_0$ is $1/(\mathrm{area}\times \mathrm{time})$. 

In Eq.~(\ref{eq:decay_nucl}), $I(t)$ is independent of the
position. This is a mean-field approximation, in which any local
reduction of the nucleation rate is smeared out over the whole
surface. Within this approximation, nucleation remains spatially
uncorrelated, and Avrami's law~[Eq.~(\ref{eq:avrami})] holds. Hence we
need to calculate the extended coverage that follows from
Eq.~(\ref{eq:decay_nucl}). Inserting $I(t)$ into
Eq.~(\ref{eq:ext_cov}) and performing the integration yields
\begin{equation}
\theta_{\mathrm{ext}}(t) = \frac{\pi I_0 v^2 t^2}{\lambda} - \frac{2\pi
  I_0 v^2 t}{\lambda^2} + \frac{2\pi I_0 v^2}{\lambda^3}
  \left(1-\mathrm{e}^{-\lambda t}\right). \label{eq:ext_cov_decay}
\end{equation}
In the limit of $\lambda t \gg 1$, the second and the third term are
of the order of $\lambda t$ and $\lambda^2 t^2$ smaller than the first
term, and we recover the well-known result for instantaneous
nucleation, Eq.~(\ref{eq:ext_cov_inst}), with $I = I_0/\lambda$. In
the limit $\lambda t \ll 1$, the exponential can be expanded up to
third order, which is the lowest non-vanishing order, and we
recover the well-known result for continuous nucleation,
Eq.~(\ref{eq:ext_cov_cont}), with $I=I_0$. 

We note that the exponential decay of the extended nucleation rate is
not derived from a physical model, but rather simply a mathematically
convenient assumption. Moreover, in real systems as
well as in our microscopic model system, the reduction of the
nucleation rate is thought to be \textit{local} in exclusion zones
around existing clusters. This would violate
the assumption that nucleation events are uncorrelated. Such
correlations are ignored by the theory presented. A theory that
includes these effects could not make use of Avrami's law.

A fit of experimental or simulation results for the relaxation
function to Eqs.~(\ref{eq:avrami}) and (\ref{eq:ext_cov_decay}) can
yield estimates for $\lambda$ and for $I_0v^2$, but no separate
estimates for $I_0$ and $v$.

\section{Model and algorithm}
\label{par:model}

\subsection{Model}
\label{par:model:model}
We model the electrodeposition process using a square lattice-gas
model with attractive nearest-neighbor interactions. Since the
subcritical fluctuations of the adsorbate are at quasi-equilibrium, a
grand-canonical Hamiltonian is used:
\begin{equation}
\mathcal{H} = -\Phi \sum_{\langle i,j\rangle}c_i(t)c_j(t) - \mu \sum_i
c_i(t), \label{eq:hamiltonian}
\end{equation}
where $\Phi$ is the lateral interaction constant and $\mu$ the
electrochemical potential of the adsorbing ion in the
solution~\footnote{For a spin $s_i=\pm 1$ Ising model with $z$ nearest
neighbors (here, $z=4$) and $\mathcal{N}$ sites, coupling constant $J$
and external magnetic field $H$, the mapping between the models is
$\mathcal{H}_I=\mathcal{H}_{LG} + (\mathcal{N}/2)\,\left[\mu -
(\mu_0/2)\right]$, $c_i=(s_i+1)/2$, $\Phi=4J$, $\mu-\mu_0=2H$, and
$\mu_0=-2zJ$~\cite{RikvoldBook94}.}. Mass-transfer limitations from the solution to the
surface are neglected. The occupation variables $c_i$ take the value 0
when site~$i$ is empty, and 1 when it is occupied.

We perform simulations for an $L \times L= 256 \times 256$ lattice
with periodic boundary conditions, using for all simulations the
interaction constant $\Phi = 4$ and the temperature $T =
0.8\,T_{\mathrm{c}}\approx 0.454\,\Phi=1.815$, where $T_{\mathrm{c}}$
is the critical temperature of the Ising model for the given
parameters~\cite{Onsager44}. Energy and
temperature units are chosen such that Boltzmann's constant
$k_{\mathrm{B}}=1$.

\subsection{Algorithm}
\label{par:model:calc}
The dynamics of the phase transformation are investigated using
kinetic Monte Carlo simulations~\cite{BinderBook97}. We assume that
the model dynamics can be reasonably approximated by a few stochastic
elementary processes; hence, we choose one-step Arrhenius-type
dynamics~\cite{Kang89,Fichthorn91} for the transition probabilities,
introducing a local transition barrier $\Delta$. The transition
rate~$R$ for a transition from configuration~$a$ to
configuration~$b$ is approximated by the corresponding rate for
a transition from $a$ to a fictitious transition state:
\begin{equation}
R(a\rightarrow b) = \nu_0 \exp\left(-\frac{\mathcal{H}^{\dagger}-
  \mathcal{H}(a)}{k_{\mathrm{B}}T}\right), \label{eq:arrh_rate}
\end{equation}
where $\mathcal{H}^{\dagger}$ is the energy at the transition
state. Using a symmetric ($\alpha = 1/2$) Butler-Volmer- (Br{\o}nsted-)type expression
for $\mathcal{H}^{\dagger}$~\cite{BrownBook99}:
\begin{equation}
\mathcal{H}^{\dagger}=\frac{\mathcal{H}(a)+ \mathcal{H}(b)}{2} + \Delta
, \label{eq:butler_volmer}
\end{equation}
one obtains
\begin{equation}
R(a\rightarrow b) = \nu_0 \exp\left(-\frac{\Delta}{k_{\mathrm{B}}T}\right)\,
\exp\left(-\frac{\mathcal{H}(b)-
  \mathcal{H}(a)}{2k_{\mathrm{B}}T}\right), \label{eq:trans_prob}
\end{equation}
where $\nu_0$ sets the Monte Carlo time scale and will be chosen as
unity in the following. Another choice for an Arrhenius-type
dynamic is the transition-dynamic algorithm~\cite{Ala-Nissila92,Ala-Nissila92a}. 

For each kind of elementary step that is included in our model
simulations, we specify a value for the transition
barrier~$\Delta$. In particular, these are $\Delta_{\mathrm{ad/des}}$ for
adsorption/desorption steps and $\Delta_{\mathrm{dif}}$ for diffusion
steps to a nearest-neighbor site. The former is equivalent to a
spin-flip step, the latter to a spin-exchange step in the terminology
of magnetic spin models. By varying the difference
$\Delta_{\mathrm{ad/des}}- \Delta_{\mathrm{dif}}$ we can set the ratio
between adsorption/desorption and diffusion steps. The absolute values
have no particular meaning unless we attempt to specify the
relation between physical and Monte Carlo time scales~\cite{Abou04}. We treat
$\Delta$ as merely formal and note that, according to
Eq.~(\ref{eq:butler_volmer}), $\mathcal{H}^{\dagger}$ might not be
larger than both $\mathcal{H}(a)$ and $\mathcal{H}(b)$ for a strongly
exothermic or endothermic step, viz not represent a true barrier, even
for positive $\Delta$. In the Monte Carlo algorithm chosen, transition
rates $R > 1$ pose no problem (see below). 

At high diffusion rates, standard Monte Carlo algorithms are
prohibitively slow since much simulation time is spent on the
coverage-conserving diffusion steps. It is therefore necessary to use
more advanced methods: we choose the $n$-fold way algorithm of Bortz,
Kalos and Lebowitz~\cite{Bortz75} and Gilmer~\cite{Gilmer76}. This
algorithm is rejection-free: of all possible new configurations, one
is chosen with probability $R_j/ R_{\mathrm{tot}}$, where
$R_{\mathrm{tot}} = \sum_i{R_i}$, and then accepted with probability
1. No time is spent on choosing sites and calculating energy
differences for rejected moves. After each step the simulation time is
updated by $-(1/R_{\mathrm{tot}})\ln r\,$MCSS (Monte Carlo steps per
site), where $r$ is drawn from a uniform pseudo-random distribution
between 0 and 1.  We need to keep a list of the transition rates for
all possible moves from the actual configuration. This is done for a
single-site $n$-fold way as follows. For each site~$s$ in the lattice
we compute the transition rate~$R_{s,m}$ for every possible elementary
step~$m$ according to Eq.~(\ref{eq:trans_prob}): adsorption/desorption
($R_{s,1}$; adsorption when site is empty, desorption otherwise);
exchange with nearest neighbor to the right ($R_{s,2}$; equal to zero
when both are occupied or both are empty), exchange with nearest
neighbor below ($R_{s,3}$; equal to zero when both are occupied or
both are empty). This minimizes the possible steps per site and makes
sure that every possible diffusion step on the lattice is counted
once. We then calculate the total rate at which site~$s$ performs a
move as $S_s = \sum_m{R_{s,m}}$. In a Monte Carlo step, a random
number is drawn from a uniform distribution to choose with probability
$S_s/R_{\mathrm{tot}}$ the site $s$ that performs the next move. Then,
another random number is drawn to choose with probability
$R_{s,m}/S_s$ which elementary step $m$ is performed. Finally, a third
random number is drawn to choose the time step by which the simulation
clock is updated. After that, the respective $R$ and $S$ of all sites
that are affected by the move are updated. In order to accelerate this
process, the latter are stored in a binary tree.

Recognizing that there is a finite number of configurations of a site
plus environment, each characterized by an energy that is constant
throughout the simulation, it is in principle possible to tabulate the
energy differences that are connected with each change of local
configuration (called classes). This has previously been worked out
for the square Ising lattice without spin-exchange
(diffusion)~\cite{Bortz75}, where 10 classes are to be considered for
adsorption/desorption. Here, we additionally have to define 32 classes
for the diffusion step; our $n$-fold way is hence a 42-fold way
algorithm.  The situation is further complicated by the
fact that the diffusion step calls for including a larger portion of
the lattice into the affected environment of a site. We have to
consider a total of 21 classes up to third-nearest neighbor positions
after an adsorption/desorption step and of 31 classes up to
fourth-nearest neighbor positions after a diffusion step. Details are
shown in the Appendix.

\subsection{Model calculations}
Our goal is the investigation of the influence of diffusion on the
dynamics of the first-order phase transition. To that end, we vary the
diffusion barrier $\Delta_{\mathrm{dif}}$ between 5 and 15, keeping
$\Delta_{\mathrm{ad/des}}=15$ constant. Moreover, we perform
simulations without diffusion, corresponding to
$\Delta_{\mathrm{dif}}=\infty$. We start our simulations with an empty
lattice, and at $t=0$ we set the electrochemical potential to a value between $\mu
- \mu_0 = 0.4$ and $\mu - \mu_0 = 1.6$, where $\mu_0$ is the
electrochemical potential at coexistence. For all these values of the
electrochemical potential, the phase transformation is in the
multi-droplet regime~\cite{Ramos99}, where the number of supercritical
adsorbate clusters is large. For each set of parameters, we
perform 500 to 1000 independent simulation runs. This high number is
necessary to accurately obtain the variance of the coverage. For quantities that
do not depend on higher moments, we typically take statistics from
only about 100 runs.

We sample the coverage~$\theta$ as a function of the simulation
time. Moreover, we follow the size of every cluster in the
lattice throughout the simulation until the coverage reaches $\theta =
1/2$; this enables us to get microscopic interface velocities and nucleation
rates. Later, cluster sizes are obtained using the Hoshen-Kopelman
algorithm~\cite{Hoshen76} at certain coverages. We compute the
time-dependent two-point correlation function~(see Ref.~\cite{Ramos99},
correcting for missing factors of powers of $L$)
\begin{equation}
G(\vec{r},t) = \langle c_i(t) c_j(t) \rangle - \langle \theta(t)
\rangle^2, \label{eq:corrf}
\end{equation}
where $\vec{r}=\vec{r_i}-\vec{r_j}$, as $1/\sqrt{L^2}$ times the
inverse Fourier transform of the structure 
factor
\begin{equation}
S(\vec{q},t) = \langle \hat{c}_{\vec{q}} \hat{c}_{-\vec{q}} \rangle -
L^2 \langle \theta(t) \rangle^2 \delta_{\vec{q},\vec{0}},  \label{eq:structfac}
\end{equation}
where $\delta_{\vec{q},\vec{0}}$ is the Kronecker delta function and
$\hat{c}_{\vec{q}}$ is the Fourier transform of $c_i$:
\begin{equation}
  \hat{c}_{\vec{q}}(t) = \frac{1}{\sqrt{L^2}} \sum_{i=1}^{L^2}{c_i(t)
  \exp(-\mathrm{i}\vec{q}\cdot\vec{r_i})}. \label{eq:FTofc}
\end{equation}
The angular brackets denote averages over independent runs. Due to the
use of a Fast Fourier Transform algorithm~\cite{NumRec}, this approach is
significantly faster than the direct calculation of $G(\vec{r},t)$. We
only report the circularly averaged $G(r,t)$.

\section{Results}
\label{par:results}

Our interest is in the influence of diffusion on the kinetics and
dynamics of the phase transformation. It is therefore necessary to
first characterize how the fraction of steps that are diffusion steps
depends on the model parameters. We count the diffusion steps and
divide their number by the number of total steps. Since the time step
by which the clock is updated after a successful move does not depend
systematically on the type of move performed, this simpler approach is
approximately equivalent to computing true rates of the elementary
processes. Figure~\ref{fig:diffrate} shows that the fraction of
diffusion steps changes only weakly during the phase
transformation. Hence, it is sufficient to discuss the average
fraction of diffusion steps throughout the whole simulation. In
Table~\ref{tab:diffrate}, we show that the diffusion fraction increases
strongly with decreasing diffusion barrier $\Delta_{\mathrm{dif}}$,
reflecting the exponential law, Eq.~(\ref{eq:trans_prob}), and
increases weakly with the electrochemical potential, reflecting that
diffusing particles spend on average longer time on the surface before
they desorb. With $\Delta_{\mathrm{dif}} = 5$, diffusion is the
predominant process.

The critical cluster size~$n^*$ is governed by the balance of volume
and surface free-energy terms of the
clusters~\cite{RikvoldBook94,Gunton83}; 
it hence depends on the interaction constant $\Phi$ (which
we keep constant), and on the electrochemical potential $\mu$ (which
we vary). The critical cluster size is the number of particles $n$ in
the cluster for which the free energy $F(n)$ has a maximum. We
calculate $F(n) = -k_{\mathrm{B}}T\ln{Z(n)}$ from the restricted
partition function $Z(n)$, which we obtain from enumerating lattice
animals, hence counting all possible configurations of a given cluster
size~\cite{Frank05}. The resulting critical cluster size is shown as a
function of $\mu$ in Table~\ref{tab:critcl}. As expected, it decreases
with increasing $\mu$.

\subsection{Dynamics of the phase transformation}
\label{par:results:dynamics}
We characterize the phase transformation using the relaxation
function~$\phi$~\cite{Ramos99}, which is closely related to
$1-\theta$ and decays from 1 to 0 during the adsorption. It is
defined as
\begin{equation}
\phi(t)= \frac{\theta(t)-\theta_{\mathrm{s}}}{\theta(0)-\theta_{\mathrm{s}}},
\end{equation}
where $\theta_{\mathrm{s}}$ is the stable coverage, which depends on the
electrochemical potential and is easily obtained from equilibrium
Monte Carlo simulations. We define the metastable lifetime~$\tau$ as
the time when $\phi = 1/2$.

The influence of the electrochemical potential on the kinetics of the
phase transformation is shown in Fig.~\ref{fig:relaxfc_nodif}. The
phase transformation lags behind the potential switch, but it is
markedly accelerated when the electrochemical potential, which is the
driving force for adsorption, is higher. In
Fig.~\ref{fig:relaxfc_H0.2} we show the influence of diffusion on the
kinetics for $\mu-\mu_0=0.4$. At $\Delta_{\mathrm{dif}} = 15$, the
phase transformation is only slightly faster than in the absence of
diffusion; however, at lower values of the diffusion barrier, the
acceleration is considerable. This trend is verified in
Table~\ref{tab:tau} for all electrochemical potentials and diffusion
barriers considered.

In a previous investigation~\cite{Frank05}, using Glauber dynamics, we
found that when diffusion is sufficiently fast on the time scale of
adsorption/desorption, there is a crossover from continuous to
instantaneous nucleation. We demonstrate in Fig.~\ref{fig:avrami} that
this behavior persists when introducing local barriers into the
dynamics. We normalize the time scale with the metastable lifetime;
then, when only the time scale of the phase transformation changes and
not the dynamics, the plots should collapse. This is the case for the
simulations without diffusion and with $\Delta_{\mathrm{dif}}=15$, but
for lower diffusion barriers we find increasingly strong
deviations. Figure~\ref{fig:avrami}(a) shows the plot for the model of
continuous nucleation [Eqs.~(\ref{eq:ext_cov_cont}) and
(\ref{eq:avrami})]. The deviations from linearity at early times
probably come from an overlap of the time scales of the formation of
the metastable phase (starting with a completely free surface) and of the
metastable decay. However, in the absence of diffusion, there is a
marked region where linear behavior is found. This region becomes
shorter and eventually disappears when diffusion becomes
predominant. Instead, as shown in Fig.~\ref{fig:avrami}(b), for
$\Delta_{\mathrm{dif}}=5$ the prediction for instantaneous nucleation
[Eqs.~(\ref{eq:ext_cov_inst}) and (\ref{eq:avrami})] is
fulfilled. Similar results were reported in Ref.~\cite{Frank05}.

\subsubsection{Crossover to instantaneous nucleation}
\label{par:results:crover}
On the basis of the results from the previous section, we attempt to
fit the phase transformation curves to our model for decaying nucleation
rates~(Sec.~\ref{par:theo:decaynucl}). According to Ref.~\cite{Ramos99},
when the KJMA theory is used as a coarse-grained approximation for the
microscopic Ising (or lattice-gas) model, the relaxation function can
be corrected for the fact that the simulation does not start from
the metastable phase:
\begin{equation}
\phi(t) \approx \frac{\theta_{\mathrm{ms}} -
  \theta_{\mathrm{s}}}{\theta(0)-\theta_{\mathrm{s}}} \phi_{\mathrm{KJMA}}(t). \label{eq:corr_phi}
\end{equation}

In a first step, we fit $\phi(t)$ from the simulations without
diffusion to Eqs.~(\ref{eq:ext_cov_cont}), (\ref{eq:avrami}), and
(\ref{eq:corr_phi}) and obtain estimates for the two free parameters
of the fit, $Iv^2$ and $\theta_{\mathrm{ms}}$. To do so, we need an
estimate for $\theta_{\mathrm{s}}$, which we obtain from equilibrium Metropolis
Monte Carlo simulations of the order of $10^6$ MCSS. The
thermodynamic quantity $\theta_{\mathrm{ms}}$ is only a function of
$T$ and $\mu$ and independent of the microscopic details of the
dynamics. It can therefore be used later for the simulations with
diffusion at the same potential. Strictly, due to the imperfect
separation of the timescales for the metastable decay and the
formation of the metastable phase from the empty surface, the
estimates for $\theta_{\mathrm{ms}}$ obtained in this way are compromised by a
systematic error. In particular at higher potentials they appear to be
too high, and we refrain from reporting their numerical
values. However, the error is most likely dominated by the
potential and not by the diffusion rate, and we use the same
$\theta_{\mathrm{ms}}$ for all diffusion rates at the same potential
in the following.

In a second step we fit $\phi(t)$ from all the simulations (with
and without diffusion) to Eqs.~(\ref{eq:avrami}),
(\ref{eq:ext_cov_decay}), and (\ref{eq:corr_phi}). The free parameters
of the fit are $I_0v^2$ and $\lambda$. The former contains information
about the kinetics of the phase transformation, the latter about the
decay time of the nucleation, viz the dynamics of the phase
transformation. We fit the relaxation function in the interval from
$0.5\tau$ to $\tau$, since at short times the decay to the metastable
phase affects $\phi$, and at longer times coalescence effects become
noticeable~\cite{Frank05}. By dividing the data into about 10 to 20
subsets and performing the fit independently for each subset, we
obtain the standard deviation of the fitting parameters,
and hence an estimate of the standard error.

In Table~\ref{tab:ivsq} we show the results for the kinetic parameter
$I_0v^2$. The standard error is insignificant for the number of digits
given. The values of $Iv^2$ reflect the same trends as the metastable
lifetime $\tau$. In the absence of diffusion and for
$\Delta_{\mathrm{dif}}=15$, the values are identical to those obtained
for $Iv^2$ from fits to Avrami's law for continuous nucleation. For
the values for which we expect a possible crossover to instantaneous
nucleation to be most complete ($\mu-\mu_0 = 1.6$,
$\Delta_{\mathrm{dif}}= 5$) $I_0v^2 / \lambda = 1.3\times 10^{-9}$,
in satisfactory agreement with $Iv^2 = 9.5\times 10^{-10}$ from a fit to
Avrami's law for instantaneous nucleation.

This behavior is reflected in the values of $\lambda\tau$ that we
report in Table~\ref{tab:lambdatau}. When $\lambda\tau =1$, the
nucleation rate has decayed by a factor $1/\mathrm{e}$ at the metastable
lifetime. Hence, when $\lambda\tau \approx 1$, we are in a transition
regime where the Avrami plots neither resemble instantaneous nor
continuous nucleation. In the regime of continuous nucleation,
$\lambda\tau \ll 1$; the exact numerical value is insignificant and
thus scatters. This can cause quite large relative standard
errors. When $\lambda\tau$ becomes larger, we are approaching the
regime of instantaneous nucleation. These finite values of
$\lambda\tau$ are indeed significant, as shown by the quite small
statistical errors in these cases. As the results in the previous
paragraph suggest, for the highest potential and lowest diffusion
barrier considered, the nucleation is virtually instantaneous. A value
of $\lambda\tau \approx 4$ means that the extended nucleation rate $I$
is reduced to about $I_0/50$ when the relaxation function reaches
$1/2$. We show the fits for $\mu-\mu_0 = 1.2$ in
Fig.~\ref{fig:avrami_fit_crover} to demonstrate how the model captures
the crossover from continuous to instantaneous nucleation. Apart from
deviations at early times, the fits are satisfactory. The decrease in
curvature with increasing diffusion rates is well captured.

\subsubsection{Morphological changes}
\label{par:res:morphology}
In Ref.~\cite{Frank05} we demonstrated that the crossover from
continuous to instantaneous nucleation is accompanied by changes in
the adsorbate morphology during the phase transformation. In
particular, the cluster-size distributions showed a depletion of
mid-sized clusters and an enrichment of large clusters. Here we show
that similar behavior is found using dynamics with local barriers. 

According to Sekimoto's theory for uncorrelated, continuous
nucleation~\cite{Sekimoto86,Sekimoto91}, the two-point correlation
function $G(r)$ is a monotonically decreasing function of the distance
$r$. We show the results for $\mu-\mu_0=0.4$ and
$\Delta_{\mathrm{dif}} = \infty$ in Fig.~\ref{fig:corrf}~(a). The
correlation functions indeed decay monotonically for all coverages. As
can be easily shown from the definition of $G(r)$, $G(0) =
\langle\theta\rangle - \langle\theta\rangle^2 =
\langle\theta\rangle\left(1-\langle\theta\rangle\right)$, which equals
the mean-square fluctuation of the occupation variable from its
average~\cite{Debye57}. For later use, it is convenient to normalize
the correlation function as $\tilde{G}(r)=G(r)/G(0)$, such that
$\tilde{G}(0)=1$.  The un-normalized $G(0)$ assumes the same value for
coverages that add up to 1, and is maximal for $\theta=0.5$. The
high-coverage correlation function of these pairs always decays more
slowly, showing the asymmetry of the morphology about $\theta=0.5$. It
has been shown in Ref.~\cite{Ramos99} that the agreement of Sekimoto's
theory with $G(r)$ is very good in the absence of diffusion. Only at
short length scales are there some deviations due to the in-phase
correlations which are neglected in the theory. The effect of
diffusion on $G(r)$ is shown in Fig.~\ref{fig:corrf}~(b). Here, for
$\mu-\mu_0=1.6$ and $\Delta_{\mathrm{dif}} = 5$, the correlation
function decays to negative values, and after going through a minimum
converges to 0. The negative values of $G(r)$ reflect the exclusion zones around
existing clusters, in which nucleation is diminished since aggregation
to the cluster is more likely than the formation of a new
cluster. This result is very important since it shows a violation of
the assumptions behind the KJMA theory and makes the use of Avrami's
law in the interpretation of the phase-transformation kinetics
questionable. Any quantitative errors of the nucleation rate and
interface velocity estimated on the basis of the KJMA theory are most
likely due to these diffusion-induced correlations. We note that
similar minima are also found in the two-point correlation functions
during phase separation with a conserved order
parameter~\cite{Brown99}, but not during phase ordering with a
non-conserved order parameter. As in the case studied here, negative
correlations during phase separation are mediated by diffusion. In our
simulations, the minima occur first when $\lambda\tau$ is of the order
of 1 (see Section~\ref{par:results:crover}). They are most pronounced
around $\theta=0.5$. With increasing coverage they move toward larger
$r$ since the average cluster size increases.

From the inverse of the initial slope of the normalized correlation
function $\tilde{G}(r)$ we can estimate the correlation length~$l$. As
we show in Fig.~\ref{fig:corrl}, the correlation length as a function
of the coverage is a concave function with the maximum shifted to
values larger than $\theta=0.5$. This shows that correlations are
of longer range for the vacancy clusters at coverage $1-\theta$ than for the
corresponding adatom clusters at coverage $\theta$. Generally, the
correlation length decreases with increasing $\mu$, and increases with
decreasing $\Delta_{\mathrm{dif}}$, indicating that diffusion causes
the formation of larger structures on average. This is corroborated by
an investigation of the cluster-size
distributions. Figure~\ref{fig:clustdistr} shows as a typical example
the distributions for $\mu-\mu_0 = 1.6$ at $\theta=0.3$. With rapid
diffusion, the densities for the largest clusters are enhanced at the
expense of mid-sized clusters. Similar results were found in
Ref.~\cite{Frank05}.

Some of these observations can easily be verified from inspecting
snapshots of the real-space morphology, like the ones shown in
Fig.~\ref{fig:snapshots}. The asymmetry of the morphology about
$\theta=0.5$ is clearly visible comparing the snapshots at
$\theta=0.3$ and $\theta=0.7$ for all potentials and diffusion
barriers shown. There are fewer, but larger vacancy islands at the
higher coverage than adsorbate islands at the lower coverage, and
their shape is markedly more different from circular. The average
island size at the same coverage increases with decreasing potential
and with increasing diffusion rate, causing the observed trends for
the correlation length.

\subsubsection{Separate nucleation rate and interface velocity}
Within the KJMA theory for spatially random, continuous
nucleation, Sekimoto~\cite{Sekimoto86,Sekimoto91} has derived an expression for
the two-point correlation function $\Gamma(r)$ as a function of the
nucleation rate, the interface velocity, and time. $\Gamma(r)$ can be used as
a coarse-grained approximation for the correlation function $G(r)$
from our simulations. From Eqs.~(\ref{eq:structfac}) and
(\ref{eq:FTofc}) it can be shown that
\begin{equation}
S(\vec{0}) = L^2\,\mathrm{var}(\theta) =
\sum_{i}^{L^2}G(\vec{r_i}).
\end{equation}
The last equality is the inverse Fourier transform for $\vec{q} =
\vec{0}$. Hence, fitting the simulation results for
$L^2\,\mathrm{var}(\theta)$ to Sekimoto's expression for $\Gamma$ can
yield separate estimates for the nucleation rate and the interface velocity $I$ and
$v$, respectively~\cite{Ramos99}, provided the assumptions of
Sekimoto's derivation hold. Taking into account the coarse graining
and local fluctuations, one obtains~\cite{Ramos99}
\begin{multline}
L^2\,\mathrm{var}(\theta) \approx \left[\theta_{\mathrm{ms}} -
  \theta_{\mathrm{s}}\right]^2  8\pi v^2 t^2 \phi_{\mathrm{KJMA}}^2
  \times \left[\Theta\left(I v^2 t^3\right) -\frac{1}{2}\right] \\+
  \phi_{\mathrm{KJMA}} k_{\mathrm{B}}T\chi_{\mathrm{ms}} + \left[1-
  \phi_{\mathrm{KJMA}}\right] k_{\mathrm{B}}T\chi_{\mathrm{s}}.
  \label{eq:Seki_fit} 
\end{multline}
The function
\begin{equation}
  \Theta(x) = \int_{0}^{1} y \mathrm{e}^{x\Psi(y)} \mathrm{d} y
\end{equation}
is obtained by numerical integration using
\begin{equation}
  \Psi(y) =
  \frac{2}{3}\left[\arccos(y)-2y\sqrt{1-y^2} + y^3\ln
    \left(\frac{1+\sqrt{1-y^2}}{y}\right)\right].
\end{equation}
Here, $\chi_{\mathrm{s}}$ and $\chi_{\mathrm{ms}}$ describe the local
fluctuations of the stable and the metastable phase, respectively. The
former is equivalent to the susceptibility in magnetic language and
can be obtained from equilibrium Monte Carlo
simulations~\cite{LandauBook00}. $Iv^2$ in the argument of $\Theta$ is known from
the fits to Avrami's law for decaying nucleation rate
(Section~\ref{par:results:crover}). The free parameters of the fit are
the interface velocity $v$ and $\chi_{\mathrm{ms}}$. Knowledge of $v$ and
$Iv^2$ enables the calculation of $I$; we do not report the results
for $\chi_{\mathrm{ms}}$.

The model fits the data very well for low supersaturation and high
diffusion barriers. The quality of the fits decreases a little with
increasing supersaturation. For $\Delta_{\mathrm{dif}}=5$, though,
Sekimoto's model is not capable of describing the data. Obviously,
diffusion causes a violation of its underlying assumptions. In
particular, nucleation is no longer spatially random. We therefore
exclude all simulations which show minima in the two-point correlation
function $G(r)$ from this analysis. Results for the nucleation rate and
the interface velocity are shown in Tables~\ref{tab:nucrate} and
\ref{tab:growthrate}, respectively. The nucleation rate grows strongly
with the supersaturation --- by more than two orders of magnitude from
$\mu-\mu_0 = 0.4$ to $\mu-\mu_0=1.6$. The interface velocity increases in the
same sense, but less dramatically, by less than an order of
magnitude. Due to the limitations of the model, no statement can be
made about the trend with (noticeable) diffusion rates. Hence, and also
for comparison with the data without diffusion, we will in the
following use information from the microscopic details of the simulations.

\subsection{Microscopic rates}
\label{par:results:micro}
To obtain microscopic nucleation rates and interface velocities, we follow the time
development of the size (number of particles) $n$ of every cluster
during a simulation run. An adsorption step can cause the coalescence
of two or more clusters. A desorption step can cause the dissociation
of a cluster; one needs to check every time if the possible fragments
are still connected. The algorithm can be accelerated by realizing that
the maximum number of fragments is the number of nearest neighbors of
the desorbed particle. Moreover, it is faster to look for connections
in the vicinity of the desorbed particle first. 

We count coalescence and dissociation events. The evaluation of
nucleation events is described in
Section~\ref{par:res:micro:nucl}. Diffusion is treated as an adsorption
step followed by a desorption step. Beyond the percolation threshold
the algorithm is prohibitively slow, and we use it only up to
$\theta=0.5$.

\subsubsection{Interface velocity}
The KJMA theory in the present form makes use of the common assumption
that the radius of a circular cluster grows linearly with time
(Allen-Cahn approximation~\cite{RikvoldBook94,Sekimoto86,Sekimoto91}).
This assumption is expected to be very good for cluster sizes much
larger than the critical cluster, but it ignores effects of the
surface tension and of the local environment (the diffusion
field). During the phase transformation of the (Ising) square
lattice-gas model in the multi-droplet regime the clusters are not
perfectly circular~\cite{Ramos99,Frank05}, but one can define an
effective radius $r=\sqrt{n/\pi}$. We compute at $t_1$ the microscopic
interface velocity for every supercritical cluster~$i$ which has been
born at $t_{0,i}$, as $v_i=(r_i(t_1)-r_i(t_{0,i}))/(t_1-t_{0,i})$,
while excluding very recently nucleated clusters ($t_{0,i} \leq 0.75
t_1$). Here, $t_1$ is the first time the coverage in the actual
simulation runs reaches $\theta=0.1$, thus minimizing the effect of
coalescence (see below). The resulting distribution of microscopic
interface velocities for each set of simulation parameters is roughly
bell-shaped and yields the mean and
standard deviations given in Table~\ref{tab:growthrate}. The mean
microscopic interface velocities increase somewhat with increasing
supersaturation, and in quite a pronounced manner with the diffusion
rate. Where data are available, the rates from Sekimoto's theory are
in good agreement. The standard deviations are typically about half of
the mean. For the simulations in the absence of diffusion, the
interface velocity compares quite well with the results from a dynamic
mean-field approximation for a solid-on-solid interface under the same
microscopic dynamics~\cite{Buendia05u}. This is also shown in
Table~\ref{tab:growthrate}.

\subsubsection{Nucleation and coalescence rates}
\label{par:res:micro:nucl}
In classical nucleation theory, for a sufficiently low supersaturation
and for a large critical nucleus, nucleation kinetics can be described
by the Fokker-Planck equation, as the superposition of a drift and a
diffusional motion in particle-size space~\cite{Smelyanskiy99}. The
former is driven by the derivative of the cluster free energy~$F(n)$ with
respect to the cluster size $n$, and the latter is a stochastic
fluctuation of the particle size.

A nucleation event happens when a cluster grows to a size larger than
a certain cutoff~$n_{\mathrm{c}}$. Likewise, a supercritical cluster
dissolves when it shrinks back to a size less than or equal to
$n_{\mathrm{c}}$. When one supercritical cluster splits into two
supercritical clusters, this is not counted as nucleation, and when two
supercritical clusters coalesce, this is not counted as dissolution of
the vanishing cluster. At the top of the free-energy barrier, when
$n_{\mathrm{c}} = n^*$, these processes are dominated by the
diffusional motion in $n$-space. Far away from the barrier, this
diffusional motion can be neglected. The diffusion length in $n$-space
is~\cite{Smelyanskiy99}
\begin{equation}
l_{\mathrm{n}} = \left(\frac{D}{-\left(\frac{\mathrm{d}^2 F(n)}{\mathrm{d} n^2}\right)_{n=n^*}}\right)^{1/2},
\end{equation}
where $D=k_{\mathrm{B}}T$ and $F(n)$ is the free energy of a cluster
of size $n$.  From a parabolic fit of the barrier region
of $F(n)$ for $\mu-\mu_0= 0.4$ this yields $l_{\mathrm{n}}=11$. In
Fig.~\ref{fig:nucl_cutoff} we compare the raw nucleation and
dissolution rates for $n_{\mathrm{c}} = n^* = 18$ and for
$n_{\mathrm{c}} = n^* + l_{\mathrm{n}} = 29$. The difference between
the raw nucleation and dissolution rates
yields the net nucleation rate. Choosing the larger cutoff, the
diffusional motion is somewhat reduced, but not eliminated, and the
net nucleation rate takes longer to reach its plateau value.
We thus use $n_{\mathrm{c}}=n^*$ in the
following. We further note that the conditions for the validity of a
Fokker-Planck equation are not strictly fulfilled with the model
parameters used.

The nucleation rates obtained with this method are real nucleation
rates, $I_{\mathrm{r}}$. Figure~\ref{fig:nucl_rate} shows the net
nucleation rate for $\mu -\mu_0 = 1.2$ in the absence of diffusion and
for fast diffusion rates. In both cases the net real nucleation rate
rises steeply in the initial phase and goes through a maximum; it then
decreases continuously to zero. We show this for all diffusion rates
investigated at this supersaturation in
Fig.~\ref{fig:nucl_rate_tau}, where the time scale is normalized to
the metastable lifetime, separating changes in the dynamics from a
mere acceleration of the phase transformation. The decrease is much
faster when diffusion rates are high. This gives an \textit{a
posteriori} justification for using a decaying extended nucleation
rate in the analysis of the relaxation function at least on a
qualitative level. However, it is quite noticeable that even in the
absence of diffusion the real nucleation rate reaches zero when only
about half of the surface is covered. This is clearly in contradiction
to the assumed constant extended nucleation rate in KJMA theory (which
would result in a drop of the real nucleation rate to half its
maximum value at this coverage).

Figure~\ref{fig:nucl_rate} also shows the net coalescence rate, which
is the difference between the rates at which supercritical clusters
vanish by coalescence and appear by dissociation. For fast diffusion
rates it increases more strongly in the initial phase, but saturates
earlier and at lower values than in the absence of diffusion. We show
this for all diffusion rates investigated at this supersaturation in
Fig.~\ref{fig:assoc_rate_tau}, where the time scale is normalized to
the metastable lifetime. We believe that this pattern of the
coalescence rate for fast diffusion is related to the changes in
morphology~(Section~\ref{par:res:morphology}). The initially more
effective coalescence results in a reduction of the density of
mid-sized clusters and an increase of the correlation length, such
that later on fewer clusters are prone to coalesce in the more strongly
ordered adsorbate layer. The difference between the net nucleation and
coalescence rates is the net rate at which supercritical clusters are
formed. Its integral over time perfectly reproduces the time-dependent
densities of supercritical clusters, corroborating the accuracy of our
cluster-counting method and analysis.

In the absence of diffusion, the decrease of the real nucleation rate
comes from the reduction of the free surface area. Its maximum value
at low coverage thus gives an estimate of the extended nucleation
rate~$I$ in the KJMA theory. The results are shown in
Table~\ref{tab:nucrate}. They are of the same order as the results
from Sekimoto's theory, though systematically a little higher.

When the microscopic net nucleation rate reaches its maximum, the net
coalescence rates are still close to zero. This implies that the real
nucleation rates at early times so obtained should be a good estimate
for the extended nucleation rate, since the microscopic nucleation
rate is reasonably well separated from coalescence processes. Also,
the microscopic nucleation rate should be very close to the time
derivative of the density of supercritical clusters in the initial
period of the phase transformation. This is indeed borne out by our
results which are shown in Table~\ref{tab:nucrate}.

\paragraph{Simulations of the metastable phase}
Further insight into the nature of the nucleation process is gained by
simulations in which we suppress the decay of the metastable phase and
the adsorption of supercritical clusters. We start the simulations
with an empty surface and at time $t=0$ we switch on a supersaturation
favorable for adsorption. We follow the sizes of all clusters at every
Monte Carlo step. Whenever a step produces a supercritical cluster,
this whole cluster is desorbed and the event counted as a nucleation
event. The histogram for the time between two nucleation events,
$t_{\mathrm{nucl}}$, is shown in Fig.~\ref{fig:nucl_times} as a log-lin plot
for $\mu-\mu_0=1.2$; the plots for the other supersaturations look
similar. The probability distribution of $t_{\mathrm{nucl}}$ decays
exponentially, as expected for a Poisson process. A further indication
is the fact that the standard deviation of $t_{\mathrm{nucl}}$ equals
its mean $\langle t_{\mathrm{nucl}}\rangle$ (see Table~\ref{tab:nucrate}). It must
be noted that, since there are never any stable clusters present on
the surface the effects of spatial correlation on the nucleation rate
are neglected. The nucleation rates obtained from the reciprocal of
$L^2\,\langle t_{\mathrm{nucl}}\rangle$ are about a factor two larger than the microscopic
rates obtained from the simulations of the phase transformation. This
corresponds to a transmission factor of about $1/2$ for the
diffusional motion across the nucleation barrier in $n$-space. For
fast diffusion rates the discrepancy is a little larger, which might
be an effect of the absence of exclusion zones in which no nucleation
can take place in these ``simulations of the metastable phase''. We emphasize that
no real metastable phase is produced by this method since the complete
desorption of supercritical clusters disturbs the quasi-equilibrium.

\subsection{Accuracy of the continuum description}
\label{par:results:continuum}
The time-dependent law for the extended nucleation rate that we used
for the analysis of the phase-transformation dynamics is a mean-field
approximation whose accuracy is difficult to predict (see
Section~\ref{par:theo:decaynucl}). Our simulations have shown a strong
decrease of the real nucleation rates in the presence of fast
diffusion. However, not even in the absence of diffusion does the
real nucleation rate follow exactly the assumptions of the theory
(Section~\ref{par:res:micro:nucl}). Moreover, the two-point
correlation functions of the adsorbate phase indicate that the
condition of spatially random nucleation is violated when diffusion
rates become noticeable (Section~\ref{par:res:morphology}). A check of
the validity of Avrami's law applying an exponentially decaying
extended nucleation rate to describe the phase transformation in the
present microscopic model, is to compare the kinetic parameter
$Iv^2_{\mathrm{model}}$ from the fit of the relaxation function to the
theory with $Iv^2_{\mathrm{micro}}$ calculated from the microscopic
rates $I$ and $v$. We assume here that the latter represents the
correct value. This is shown in Fig.~\ref{fig:micro_ivsq}.  

In the absence of diffusion and for low diffusion rates the agreement
is very good. When the diffusion rate increases, we find systematic
deviations: the result from the fit is increasingly too low. Still the
fits yield the correct order of magnitude for $Iv^2$. Hence, the
agreement between the KJMA theory for exponentially decaying
nucleation rates and the microscopic results is quite satisfactory,
keeping in mind that the former is only a coarse-grained continuum
approximation of our microscopic model system. In particular, the
incomplete separation of the time scales for the formation and the
decay of the metastable phase is a potential source for deviations and
limits the quality of the curve fitting. Introducing an additional
fitting parameter that shifts the time scale might improve the ability
of the model to fit the shape of the phase-transformation curves, but
its physical meaning would be questionable. 

In the limit of instantaneous nucleation, $I_0 / \lambda$ should equal
the density of nuclei. This can be checked using the microscopic
$I_{\mathrm{r}}$ at early times as an estimate for $I_0$ and the
maximum density of supercritical clusters as an estimate for the
density of nuclei. The results are presented in
Table~\ref{tab:nucdens}. They show quite good agreement and the
correct trend with the supersaturation, despite the deviations of
$Iv^2$ under the respective conditions. Hence, $\lambda$ is indeed
physically meaningful and not a mere model parameter.

We point out that the presented theory does not account for spatial
correlations between the nuclei that are a consequence of the
diffusion-induced exclusion zones. It is difficult to anticipate the
consequences of an appropriate extension of the theory. However, since
these correlations shift the overlap between growing extended clusters
to later times, accelerating the phase transformation at early times,
one might speculate that the discrepancies in $Iv^2$ would rather
increase. We conclude that even though some of the assumptions behind
the KJMA theory are not strictly fulfilled, it still provides a
reasonable description of the phase-transformation kinetics of the
Ising lattice gas. The effects of adsorbate diffusion can in part be
accounted for by assuming a decaying nucleation rate. This effectively
yields a crossover from continuous to instantaneous nucleation when
diffusion is the predominant process.

\subsection{Onset of the crossover}
\label{par:results:onsetcrover}
Our previous results have shown that a crossover of the nucleation
from continuous to instantaneous sets in when the fit of the relaxation
function to Eqs.~(\ref{eq:avrami}), (\ref{eq:ext_cov_decay}), and
(\ref{eq:corr_phi}) results in a parameter $\lambda \tau$ of the order
of one. Then, we find changes in the morphology of the adsorbate phase
and a faster decrease of the microscopic nucleation rate. As commonly
accepted in vacuum surface science, diffusion of adsorbed particles results
in exclusion zones around existing clusters, in which nucleation is
suppressed, since free particles rather aggregate to the cluster than
form a new nucleus. Our hypothesis is that these exclusion zones
become space-filling at early times for high diffusion
rates~\cite{Frank05}. Here we present a rough estimate of when the
crossover sets in.

The average distance between independent droplets is roughly~\cite{Ramos99}
\begin{equation}
R_0 \approx v \tau.
\end{equation}
From the competition between desorption and diffusion, a monomer after
the time it adsorbs performs on average
\begin{equation}
N(\infty) = \frac{p_{\mathrm{dif}}}{p_{\mathrm{des}}}
\end{equation}
steps.
Here, from Eq.~(\ref{eq:trans_prob}) $p_{\mathrm{dif}} = \nu_0
\exp(-\Delta_{\mathrm{dif}}/k_{\mathrm{B}}T)$ is the diffusion
probability of a free monomer, and $p_{\mathrm{des}} = \nu
\exp(-\Delta_{\mathrm{ad/des}}/k_{\mathrm{B}}T)\exp(-\mu/2\,k_{\mathrm{B}}T)$
with $\mu = (\mu-\mu_0) - 2\Phi$ is its desorption probability. In a
two-dimensional random walk the particle travels a distance
$\sqrt{N(\infty)}$. We compare $\sqrt{N(\infty)}$ and $R_0$ in
Table~\ref{tab:onset}. For $\Delta_{\mathrm{dif}}=15$ the monomer performs
on average less than one step. The crossover to instantaneous
nucleation is complete when the diffusion length of a monomer is about
half the average cluster distance. Despite the crudeness of the
approximation it gives a satisfactory explanation for the onset of the
crossover. It correctly reproduces the trends with supersaturation and
diffusion barrier.

\section{Summary and Conclusions}
\label{par:summary}
We have performed kinetic Monte Carlo simulations of the adsorption
dynamics of a monolayer in a square lattice-gas model using Arrhenius
dynamics with local barriers. The model Hamiltonian is
grand-canonical, meaning that local adsorption/desorption equilibrium
is approximately established. This makes it suitable for submonolayer
electrochemical deposition under control of the potential difference
between solution and electrode and when mass-transport limitations to
the surface are insignificant. The elementary steps included in the
model are adsorption, desorption, and diffusion; our main interest is
in the influence of the latter on the phase-transformation
dynamics. The diffusion rates can be controlled by the barrier height
for the diffusion steps. We have obtained results for the
phase-transformation kinetics, the adsorbate morphology, and the
microscopic rates of nucleation, cluster growth (interface velocity),
and coalescence. For our model calculations we have used an extension
of Bortz et al.'s $n$-fold way algorithm that contains both
adsorption (spin-flip) and diffusion (spin-exchange), and whose details
we present in this paper.

The nucleation rate and the interface velocity and consequently the
phase-transformation kinetics are accelerated by both an increase of
the supersaturation (the driving force) and the diffusion rate. The
higher the diffusion rate, the more likely it is that a walker
aggregates with sub- or supercritical clusters to participate in
nucleation or growth, respectively. 

Around supercritical clusters, diffusion causes a local reduction of
the nucleation rate in a zone with the extension of the monomer
diffusion length, since it is more likely that monomers reach the
existing cluster than that they create a new one. This causes a
reduction of the cluster density and an increase of the correlation
length. Moreover, diffusion changes the spatial correlation between the
nuclei, which is no longer random, and creates minima in the two-point
correlation functions.

For a theoretical description of the phase transformation we use the
Kolmogorov-Johnson-Mehl-Avrami (KJMA) theory as a coarse-grained
approximation to our microscopic model system. The theory is based on
the assumption that nucleation is spatially random. Using an
appropriate law for the time dependence of the nucleation rate, one can
obtain an estimate for $Iv^2$ from a fit of the time dependence of the
relaxation function of the metastable phase, but no separate estimates
for the nucleation rate~$I$ and the interface velocity~$v$. In the absence of
diffusion, this gives very accurate results assuming a constant
extended nucleation rate~$I$ (continuous nucleation), even though
our microscopic results suggest that the real nucleation rate
decreases somewhat too fast with time for this assumption to be
strictly fulfilled. With diffusion, though, the exclusion zones around
supercritical clusters grow with the cluster radius, such that the
real nucleation rate diminishes more strongly with time than expected
from the decrease of the free surface fraction (and than in the case
without diffusion). Assuming an extended nucleation rate that
decreases exponentially with time from an initial value $I_0$, we can
extend the analysis to this case and still extract an estimate for
$I_0v^2$. This estimate is less accurate, but still correct by order
of magnitude. We emphasize that correlation effects, which are also
important in these cases, are neglected, since we use Avrami's
law. When diffusion is predominant, the exclusion zones become
space-filling at early times of the phase transformation, and the
dynamics resemble instantaneous nucleation. A fit to the
respective form of Avrami's law would, however, not yield an estimate
for $I_0v^2$. Instead, $I_0v^2/\lambda$ is obtained from this
analysis, where $\lambda$ is the time constant of the decay of
$I$. $I_0 / \lambda$ is the density of nuclei. We emphasize that
nucleation is still homogeneous. Hence, in the electrodeposition of a
two-dimensional film that proceeds by a first-order phase transition,
instantaneous nucleation need not imply heterogeneous nucleation at
defects. Also, intermediate cases are possible that would lead to
non-linear Avrami plots and non-integer Avrami exponents. A strong
indication for our model to be a good description of a given experimental
system could be found in the two-point correlation functions, which
develop minima for an Avrami exponent less then 3. They can be
obtained from scattering experiments~\cite{Mitchell01} or by analysis
of microscopic images. 

In the absence of diffusion, it is possible to obtain separate
estimates for $I$ and $v$ from the variance of the coverage
\textit{vs}. time using Sekimoto's theory for the two-point
correlation function. This theory is based on spatially random
nucleation as well, and the analysis breaks down when diffusion rates
become noticeable and a minimum in the correlation function appears.

The time a walker spends on the surface before it desorbs increases
with higher supersaturation, such that in this case the influence of
diffusion on the dynamics of the phase transformation at the same rate
is stronger. Hence, the crossover to instantaneous nucleation is more
complete for the same diffusion barrier when the supersaturation is
greater.

In a previous work~\cite{Frank05} we have presented kinetic Monte
Carlo simulations of the same model using Glauber dynamics, but
without varying the supersaturation, and performing a less stringent
analysis of the data. Still, we found from the time-dependence of the
relaxation function qualitatively the same crossover to instantaneous
nucleation as in the present investigation. In various studies for
solid-on-solid (SOS) and Ising models, a comparison of different
transition probabilities in kinetic Monte Carlo dynamics has shown
that the stochastic dynamics can have a strong influence on the
microscopic properties of the evolving system, namely interfacial
structure, nucleation rates, and interface velocities (see, e.g.,
Refs.~\cite{Park04,Buendia04,Buendia05u}). Distinctions are made
between so-called ``soft'' dynamics, in which the field
(supersaturation) and coupling (interaction) constant terms in the
transition probability factorize, and ``hard'' dynamics, in which they
do not. The dynamics in the present work are soft, while in
Ref.~\cite{Frank05} they are hard. Another difference is that here we
use local barriers in contrast to Ref.~\cite{Frank05}. In both
investigations, the stochastic dynamics contain conserving and
non-conserving steps with respect to the order parameter (the
coverage). This paper does not aim at a systematic investigation of
the influence of the stochastic dynamics; in fact the diffusion is
treated in such different ways that a quantitative comparison is
difficult. It is, however, possible to conclude that the decrease in
local nucleation rates around clusters and the concomitant
development of spatial correlations translate into qualitatively
similar changes in the phase-transformation dynamics.

\ack{
This work was supported by the U.S.\ National Science Foundation under Grant
No.\ DMR-0240078, and by Florida State University through its Center
for Materials Research and Technology and its School of Computational
Science. S.F.'s work in Ulm is supported by the Deutsche
Forschungsgemeinschaft under Grant No. HA 5159/1-1}

\renewcommand{\theequation}{A-\arabic{equation}}
\setcounter{equation}{0}  
\section*{Appendix}  
Here we introduce our single-site $n$-fold way scheme for the square
lattice-gas model with both adsorption/desorption and diffusion
steps. For computational convenience, we present the classification
scheme in Ising language, $s=+1$ representing a spin up (occupied
site) and $s=-1$ representing a spin down (empty site). For the sake
of consistency, we also present the energy differences in terms of the
Ising Hamiltonian. Within Ising terminology, adsorption/desorption is
equivalent to spin flip and diffusion to spin exchange. The former
involves only one particle, while the latter involves a pair. Hence,
care must be taken that diffusion steps are not counted twice when we
set up our list of all elementary steps in the lattice. For each
particle, only spin exchange with two of the four nearest neighbors in
orthogonal directions (here: right and below, respectively) is to be
considered. The spin exchange with the nearest neighbors in the
opposite directions (left and above, respectively) is included in
the neighbors' respective set of moves. In lattice-gas terms, instead of the
diffusion of a particle to the left or upwards, we consider the
diffusion of a vacancy to the right or downwards, starting from the
respective nearest-neighbor position. We use the following labels for
the three elementary steps associated with each lattice point: $m=1$
(adsorption/desorption), $m=2$ (diffusion to [spin exchange with] the
nearest-neighbor position to the right, and $m=3$ (diffusion to [spin
exchange with] the nearest-neighbor position below). We call the classes
for the adsorption/desorption steps particle classes and for the
diffusion steps pair classes. For the particle classes, we need to
consider the value (spin) $s_i$ of the central particle C and of its
four nearest neighbors, $s_k$. For the pair classes, we need to
consider the spins $s_i$ and $s_j$ of two neighboring particles, C and
C', respectively. C and C' form the pair S. Moreover, we consider the
spins $s_k$ of \textit{all} nearest neighbors of C, and the spins
$s_l$ of \textit{all} the nearest neighbors of C'. The central
particle C is the one to which the respective pair class belongs. C'
is called the exchange partner. 

For the particle classes $c_{\mathrm{a/d}}$, the scheme is shown in
Table~\ref{tab:part_class}~\cite{Bortz75}. From it follows the simple
relation
\begin{equation}
c_{\mathrm{a/d}}= \frac{1}{2}\left(11 - 5 s_i - \sum_k{s_k}\right).
\end{equation}
The difference of the configurational energy resulting from the respective
adsorption/desorption step is obtained from
\begin{equation}
\mathcal{H}(b) - \mathcal{H}(a) = \Delta{\mathcal{H}} = 2 J s_i
\sum_k{s_k} + 2 H s_i.
\end{equation}
For the pair classes~$c_{\mathrm{dif}}$, only the cases where $s_j
\neq s_i$ need to be considered; the scheme is shown in
Table~\ref{tab:pair_class}. It follows the simple relation
\begin{equation}
c_{\mathrm{dif}}= -\frac{13}{2} s_i + 2 \sum_k{s_k} + \frac{1}{2}
\sum_l{s_l} + \frac{33}{2}.
\end{equation}
The difference of the configurational energy performing the respective
diffusion step is obtained from
\begin{equation}
\mathcal{H}(b) - \mathcal{H}(a) = \Delta{\mathcal{H}} = J\left[
  \left(s_i-s_j\right) 
\left(\sum_k{s_k} - \sum_l{s_l}\right) + 4 \right].
\end{equation}
Not only are the classes changed by moves in which the respective site
is involved, but also by moves in the neighborhood. The classes of up
to some of the 4th-nearest neighbors have to be updated after a
move. See Fig.~\ref{fig:affected} for an overview of the sites
affected by a move, and Fig.~\ref{fig:nn_coord} for the designation of
the neighboring positions.  We denote the spin of an affected site
with $t_i$ when it is the central particle X, and with $t_j$ when it is
the exchange partner X' of X.  X and X' form the pair T. The spins of
\textit{all} the nearest neighbors of X and X' are labeled $t_k$ and
$t_l$, respectively. Tables~\ref{tab:class_update_part},
\ref{tab:class_update_pair_ad} and \ref{tab:class_update_pair_dif}
summarize how the classes in the neighborhood are affected.


\newpage
\clearpage
\section*{Figure captions}

\begin{figure}[h]
\caption{Fraction of diffusion steps as a function
of the coverage during adsorption of a monolayer. Diffusion
barrier $\Delta_{\mathrm{dif}}=15$.} \label{fig:diffrate}
\end{figure}

\begin{figure}[h]
\caption{Acceleration of the phase
  transformation with increasing electrochemical
  potential. $\Delta_{\mathrm{dif}} = \infty$.} \label{fig:relaxfc_nodif}
\end{figure}

\begin{figure}[h]
\caption{Acceleration of the phase
  transformation with diffusion. $\mu - \mu_0 = 0.4$.}
  \label{fig:relaxfc_H0.2}
\end{figure}

\begin{figure}[h]
\caption{Crossover from continuous to instantaneous
  nucleation for $\mu - \mu_0 = 1.2$. (a) Logarithmic plot of the
  relaxation function \textit{vs}. the cube of the normalized time, as
  expected for continuous nucleation, Eqs.~(\ref{eq:ext_cov_cont}) and
  (\ref{eq:avrami}). (b) Same \textit{vs}. the square of the
  normalized time, as expected for instantaneous nucleation,
  Eqs.~(\ref{eq:ext_cov_inst}) and (\ref{eq:avrami}). In both cases,
  the straight line is a guide to the eye.} \label{fig:avrami}
\end{figure}

\begin{figure}[h]
\caption{Fits of the relaxation function
  to Avrami's law for exponentially decaying nucleation rate. Dotted
  lines: simulations. Continuous lines: fits. The curves are, from right to
  left, for $\Delta_{\mathrm{dif}} = \infty$, 15, 10, 5. $\mu - \mu_0
  = 1.2$.} \label{fig:avrami_fit_crover} 
\end{figure}

\begin{figure}[h]
\caption{Two-point correlation functions
  $G(r)$ as  functions of the distance $r$ for various
  coverages. (a) $\mu-\mu_0=0.4$, $\Delta_{\mathrm{dif}}=\infty$. (b)
  $\mu-\mu_0=1.6$, $\Delta_{\mathrm{dif}}=5$. Note the different scale
  along the horizontal axis in the two parts.} \label{fig:corrf}
\end{figure}

\begin{figure}[h]
\caption{Correlation length~$l$ \textit{vs}. $\theta$, estimated
  from the inverse of the initial slope of the normalized correlation function
  $\tilde{G}(r)$. Thin lines: $\mu-\mu_0=0.4$. Bold lines:
  $\mu-\mu_0=1.6$.} \label{fig:corrl} 
\end{figure}

\begin{figure}[h]
\caption{Cluster-size distribution for
  $\mu-\mu_0=1.6$ at coverage $\theta=0.3$. The vertical line marks
  the critical cluster size $n^*$.} \label{fig:clustdistr} 
\end{figure}

\begin{figure}[h]
\caption{Snapshots of the adsorbate phase during
  the phase transformation. From left to right: $\mu - \mu_0
  = 0.4$, $\Delta_{\mathrm{dif}} = \infty$; $\mu - \mu_0
  = 0.4$, $\Delta_{\mathrm{dif}} = 5$; $\mu - \mu_0
  = 1.6$, $\Delta_{\mathrm{dif}} = \infty$; $\mu - \mu_0
  = 1.6$, $\Delta_{\mathrm{dif}} = 5$. Upper row: $\theta = 0.3$. Lower
  row: $\theta=0.7$.} \label{fig:snapshots} 
\end{figure}

\begin{figure}[h]
\caption{Nucleation and dissolution rates
  and net nucleation rate for a cutoff size equal to the critical cluster size
  (thin lines) and one diffusion length in $n$-space larger (bold
  lines). $\mu-\mu_0=0.4$, $\Delta_{\mathrm{dif}}=15$.}
  \label{fig:nucl_cutoff}  
\end{figure}

\begin{figure}[h]
\caption{Net nucleation and coalescence rates
  and net cluster formation rate. (a) $\Delta_{\mathrm{dif}} =
  \infty$. (b) $\Delta_{\mathrm{dif}} = 5$.  In both cases,
  $\mu-\mu_0=1.2$. Note the different scales in the two parts.}  \label{fig:nucl_rate} 
\end{figure}

\begin{figure}[h]
\caption{Net nucleation rate for various
  diffusion barriers. $\mu-\mu_0=1.2$.}  \label{fig:nucl_rate_tau} 
\end{figure}

\begin{figure}[h]
\caption{Net coalescence rate for various
  diffusion barriers. $\mu-\mu_0=1.2$.}  \label{fig:assoc_rate_tau}  
\end{figure}

\begin{figure}[h]
\caption{Probability distribution of the time
  between two nucleation events in the simulations of the metastable
  phase. $\mu-\mu_0=1.2$.}  \label{fig:nucl_times} 
\end{figure}

\begin{figure}[h]
\caption{The kinetic parameter $Iv^2$ from
  the microscopic rates $I$ and $v$ and from the fit of the relaxation
  function to the theory.}  \label{fig:micro_ivsq} 
\end{figure}

\renewcommand{\thefigure}{A-\arabic{figure}}
\setcounter{figure}{0}  

\begin{figure}[h]
\caption{(Color online) An overview of the affected classes of
  sites by a move of the central particle. (a) particle classes
  affected by adsorption/desorption: central site (red) and nearest
  neighbors (green). (b) particle classes affected by diffusion: the
  switching pair (red) and its nearest neighbors (green). (c) pair
  classes affected by adsorption/desorption: the pairs that contain
  the central particle (red) and those that contain one of its nearest
  neighbors (blue). (d) pair classes affected by diffusion: the
  diffusing pair (yellow), pairs that contain the central particle
  (red), pairs that contain the exchange partner (brown), pairs that
  contain nearest neighbors of the central particle (blue), of the
  exchange partner
  (pink), and of both (green). Bold arrows: the pair starts at the
  respective particle or nearest-neighbor site; thin arrows: the pair
  points to the respective site.} \label{fig:affected} 
\end{figure}

\begin{figure}[h]
\caption{Designation of the affected nearest (NN),
next-nearest (NNN), 3rd-nearest (3NN), and 4th-nearest (4NN) neighbor position
when the central site undergoes a move. The numbers are the labels
used in Tables~\ref{tab:class_update_part},
\ref{tab:class_update_pair_ad}, and \ref{tab:class_update_pair_dif}.}
\label{fig:nn_coord} 
\end{figure}

\clearpage
\newpage
\section*{Tables}

\begin{table}[h]
\begin{center}
\caption{The average fraction of diffusion steps
  throughout the phase transformation.} \label{tab:diffrate}
\begin{tabular}{llll}
$\mu-\mu_0$   &$\Delta_{\mathrm{dif}}=15$&$\Delta_{\mathrm{dif}}=10$&$\Delta_{\mathrm{dif}}=5$\\
\hline
$0.4$&0.221&0.817&0.986\\  
$0.8$&0.239&0.830&0.987\\  
$1.2$&0.247&0.834&0.987\\  
$1.6$&0.250&0.834&0.986\\  
\end{tabular}
\end{center}
\end{table}

\begin{table}[h]
\begin{center}
\caption{The critical cluster size $n^*$ as a function of
  the electrochemical potential.} \label{tab:critcl}
\begin{tabular}{ll}
$\mu-\mu_0$   &$n^*$\\
\hline
$0.4$&18\\
$0.8$&7\\
$1.2$&4\\
$1.6$&3\\
\end{tabular}
\end{center}
\end{table}

\begin{table}[h]
\begin{center}
\caption{The metastable lifetime~$\tau$ in MCSS. $\tau$
  is the time when the relaxation function $\phi=1/2$.} \label{tab:tau}
\begin{tabular}{lllll}
$\mu-\mu_0$ & $\Delta_{\mathrm{dif}}=\infty$ &
  $\Delta_{\mathrm{dif}}=15$ & $\Delta_{\mathrm{dif}}=10$ &
  $\Delta_{\mathrm{dif}}=5$\\
\hline
$0.4$&270000 &260000 &216000 &137000\\  
$0.8$&59000 &56800 &49200 &38800\\  
$1.2$&27500 &26800 &24100 &21300\\  
$1.6$&16500 &16200 &15150 &14300\\  
\end{tabular}
\end{center}
\end{table}

\begin{table}[h]
\begin{center}
\caption{The kinetic parameter~$I_0v^2$ in
  MCSS$^{-3}$, from a fit of the relaxation function $\phi$ to
  Avrami's law with an exponentially decaying nucleation rate.}
  \label{tab:ivsq} 
\begin{tabular}{lllll}
$\mu-\mu_0$ & $\Delta_{\mathrm{dif}}=\infty$ & $\Delta_{\mathrm{dif}}=15$ & $\Delta_{\mathrm{dif}}=10$ & $\Delta_{\mathrm{dif}}=5$ \\
\hline
$0.4$ & $3.0\times 10^{-17}$ & $3.3\times 10^{-17}$ & $5.9\times 10^{-17}$ & $3.3\times 10^{-16}$ \\  
$0.8$ & $2.7\times 10^{-15}$ & $3.0\times 10^{-15}$ & $5.0\times 10^{-15}$ & $1.9\times 10^{-14}$ \\  
$1.2$ & $2.6\times 10^{-14}$ & $2.8\times 10^{-14}$ & $4.5\times 10^{-14}$ & $1.2\times 10^{-13}$ \\  
$1.6$ & $1.2\times 10^{-13}$ & $1.2\times 10^{-13}$ & $1.8\times 10^{-13}$ & $4.0\times 10^{-13}$ \\  
\end{tabular}
\end{center}
\end{table}

\begin{table}[h]
\begin{center}
\caption{The parameter~$\lambda\tau$, from a fit of
  the relaxation function $\phi$ to Avrami's law with an exponentially
  decaying nucleation rate. $\lambda\tau$ indicates how much the
  extended nucleation rate $I$ has decayed at the metastable
  lifetime. Standard errors are given in brackets below the values.}
  \label{tab:lambdatau} 
\begin{tabular}{lllll}
$\mu-\mu_0$ & $\Delta_{\mathrm{dif}}=\infty$ &$\Delta_{\mathrm{dif}}=15$ & 
  $\Delta_{\mathrm{dif}}=10$ &$\Delta_{\mathrm{dif}}=5$\\
\hline
$0.4$ & $6.7\times 10^{-3}$   & $1.1\times 10^{-4}$ & $2\times 10^{-2}$   & $1.70$   \\  
      & $(6.6\times 10^{-3})$ & $(1\times 10^{-5})$ & $(1\times 10^{-2})$ & $(0.04)$ \\
$0.8$ & $8.3\times 10^{-5}$   & $7\times 10^{-5}$   & $0.36$              & $3.52$   \\
      & $(6\times 10^{-6})$   & $(1\times 10^{-5})$ & $(0.02)$            & $(0.03)$ \\
$1.2$ & $5.8\times 10^{-5}$   & $3\times 10^{-2}$   & $0.65$              & $4.13$   \\
      & $(8\times 10^{-6})$   & $(2\times 10^{-2})$ & $(0.01)$            & $(0.02)$ \\
$1.6$ & $6.9\times 10^{-5}$   & $1.5\times 10^{-4}$ & $0.79$              & $4.20$   \\
      & $(8\times 10^{-6})$   & $(3\times 10^{-5})$ & $(0.01)$            & $(0.01)$ \\
\end{tabular}
\end{center}
\end{table}

\begin{table}[h]
\begin{center}
\caption{Nucleation rate per site and
  MCSS. Extended nucleation rate $I$ from
  fit of $L^2\,\mathrm{var}(\theta)$ to Sekimoto's theory; real
  nucleation rate $I_{\mathrm{r}}$ from the
  maximum of the time-dependent net nucleation rate obtained from cluster
  counting; $I_{\mathrm{r}}$ from the initial slope of the density of supercritical
  clusters; and $I_{\mathrm{r}}$ from simulations of the metastable phase. For the
  simulations of the metastable phase, also the mean and standard
  deviation of the time in MCSS between nucleation events for the lattice size
  are given. For details of the methods, see the text.} \label{tab:nucrate}
\begin{tabular}{llllllll}
$\mu-\mu_0$ & $\Delta_{\mathrm{dif}}$ & Sekimoto &microscopic     & cluster density  & \multicolumn{3}{c}{metastable} \\ 
            &                         &  $I$     &$I_{\mathrm{r}}$& $I_{\mathrm{r}}$ &$I_{\mathrm{r}}$&$\langle t_{\mathrm{nucl}}\rangle$&$\sigma$ \\\hline
$0.4$ & $\infty$ & $7.2\times 10^{-9}$ & $1.4\times 10^{-8}$ & $1.2\times 10^{-8}$ & $2.8\times 10^{-8}$ & $556.9$ & $558.3$ \\  
      & $15$     & $8.3\times 10^{-9}$ & $1.3\times 10^{-8}$ & $1.1\times 10^{-8}$ & $2.9\times 10^{-8}$ & $528.3$ & $523.4$ \\ 
      & $10$     & $1.1\times 10^{-8}$ & $1.8\times 10^{-8}$ & $1.4\times 10^{-8}$ & $3.8\times 10^{-8}$ & $402.1$ & $400.6$ \\ 
      & $5$      &                     & $3.0\times 10^{-8}$ & $2.5\times 10^{-8}$ & $8.7\times 10^{-8}$ & $176.1$ & $174.8$ \\ 
$0.8$ & $\infty$ & $1.6\times 10^{-7}$ & $2.3\times 10^{-7}$ & $2.1\times 10^{-7}$ & $4.6\times 10^{-7}$ & $33.2$  & $33.2$ \\ 
      & $15$     & $2.0\times 10^{-7}$ & $2.5\times 10^{-7}$ & $2.3\times 10^{-7}$ & $4.9\times 10^{-7}$ & $31.2$  & $31.3$ \\ 
      & $10$     &                     & $3.1\times 10^{-7}$ & $2.7\times 10^{-7}$ & $6.6\times 10^{-7}$ & $23.1$  & $23.3$ \\ 
      & $5$      &                     & $4.3\times 10^{-7}$ & $3.9\times 10^{-7}$ & $1.2\times 10^{-6}$ & $12.8$  & $12.6$ \\ 
\end{tabular}
continued on next page
\end{center}
\end{table}
\addtocounter{table}{-1}
\begin{table}[h]
\begin{center}
\caption{continued}
\begin{tabular}{llllllll}
$\mu-\mu_0$ & $\Delta_{\mathrm{dif}}$ & Sekimoto & microscopic & cluster density & \multicolumn{3}{c}{metastable} \\ 
            &                         &  $I$     &$I_{\mathrm{r}}$& $I_{\mathrm{r}}$ &$I_{\mathrm{r}}$&$\langle t_{\mathrm{nucl}}\rangle$&$\sigma$ \\\hline
$1.2$ & $\infty$ & $6.4\times 10^{-7}$ & $9.7\times 10^{-7}$ & $9.1\times 10^{-7}$ & $1.8\times 10^{-6}$ & $8.5$   & $8.5$ \\ 
      & $15$     & $8.3\times 10^{-7}$ & $1.1\times 10^{-6}$ & $9.5\times 10^{-7}$ & $1.9\times 10^{-6}$ & $8.0$   & $7.9$ \\ 
      & $10$     &                     & $1.2\times 10^{-6}$ & $1.1\times 10^{-6}$ & $2.5\times 10^{-6}$ & $6.1$   & $6.1$ \\ 
      & $5$      &                     & $1.5\times 10^{-6}$ & $1.3\times 10^{-6}$ & $3.9\times 10^{-6}$ & $4.0$   & $3.9$ \\ 
$1.6$ & $\infty$ & $1.8\times 10^{-6}$ & $2.3\times 10^{-6}$ & $2.1\times 10^{-6}$ & $3.9\times 10^{-6}$ & $4.0$   & $3.9$ \\ 
      & $15$     & $1.8\times 10^{-6}$ & $2.4\times 10^{-6}$ & $2.2\times 10^{-6}$ & $4.1\times 10^{-6}$ & $3.7$   & $3.7$ \\ 
      & $10$     &                     & $2.7\times 10^{-6}$ & $2.4\times 10^{-6}$ & $5.1\times 10^{-6}$ & $3.0$   & $3.0$ \\ 
      & $5$      &                     & $2.9\times 10^{-6}$ & $2.6\times 10^{-6}$ & $7.0\times 10^{-6}$ & $2.2$   & $2.2$ \\ 
\end{tabular}
\end{center}
\end{table}

\begin{table}[h]
\begin{center}
\caption{Interface velocity~$v$ in unit length per MCSS. From fit of
  $L^2\,\mathrm{var}(\theta)$ to Sekimoto's theory; from the increase
  of the effective radius of the clusters during the initial phase of
  the simulation (standard deviation in parentheses); solid-on-solid
  interface from a dynamic mean-field
  approximation~\cite{Buendia05u}. For details of the methods, see the
  text.} \label{tab:growthrate}
\begin{tabular}{llllll}
$\mu-\mu_0$ & $\Delta_{\mathrm{dif}}$ & Sekimoto & microscopic & (standard deviation) & MFA\\ \hline
$0.4$ & $\infty$ & $6.5\times 10^{-5}$ & $5.3\times 10^{-5}$ & $(2.8\times 10^{-5})$ & $4.26\times 10^{-5}$\\  
      & $15$     & $6.4\times 10^{-5}$ & $5.5\times 10^{-5}$ & $(2.8\times 10^{-5})$ & \\ 
      & $10$     & $7.4\times 10^{-5}$ & $7.0\times 10^{-5}$ & $(3.3\times 10^{-5})$ & \\ 
      & $5$      &                     & $1.4\times 10^{-4}$ & $(5.5\times 10^{-5})$ & \\ 
$0.8$ & $\infty$ & $1.3\times 10^{-4}$ & $1.3\times 10^{-4}$ & $(8.0\times 10^{-5})$ & $8.58\times 10^{-5}$\\ 
      & $15$     & $1.2\times 10^{-4}$ & $1.3\times 10^{-4}$ & $(8.1\times 10^{-5})$ & \\ 
      & $10    $ &                     & $1.8\times 10^{-4}$ & $(9.9\times 10^{-5})$ & \\ 
      & $5$      &                     & $4.0\times 10^{-4}$ & $(1.6\times 10^{-4})$ & \\ 
$1.2$ & $\infty$ & $2.0\times 10^{-4}$ & $1.8\times 10^{-4}$ & $(1.3\times 10^{-4})$ & $1.30\times 10^{-4}$\\ 
      & $15$     & $1.8\times 10^{-4}$ & $2.0\times 10^{-4}$ & $(1.4\times 10^{-4})$ & \\ 
      & $10$     &                     & $2.9\times 10^{-4}$ & $(1.7\times 10^{-4})$ & \\ 
      & $5$      &                     & $6.7\times 10^{-4}$ & $(2.8\times 10^{-4})$ & \\ 
\end{tabular}
continued on next page
\end{center}
\end{table}
\addtocounter{table}{-1}
\begin{table}[h]
\begin{center}
\caption{continued}
\begin{tabular}{llllll}
$\mu-\mu_0$ & $\Delta_{\mathrm{dif}}$ & Sekimoto & microscopic & (standard deviation) & MFA\\ \hline
$1.6$ & $\infty$ & $2.5\times 10^{-4}$ & $2.3\times 10^{-4}$ & $(1.9\times 10^{-4})$ & $1.76\times 10^{-4}$\\ 
      & $15$     & $2.6\times 10^{-4}$ & $2.5\times 10^{-4}$ & $(1.9\times 10^{-4})$ & \\ 
      & $10$     &                     & $3.9\times 10^{-4}$ & $(2.4\times 10^{-4})$ & \\ 
      & $5$      &                     & $9.4\times 10^{-4}$ & $(3.8\times 10^{-4})$ & \\ 
\end{tabular}
\end{center}
\end{table}

\begin{table}[h]
\begin{center}
\caption{Ratio of nucleation rate $I_0$ and
  fitting parameter $\lambda$ as estimate for the density of nuclei,
  which is approximated by the maximum density of supercritical
  clusters. $\Delta_{\mathrm{dif}} = 5$.}  \label{tab:nucdens}
\begin{tabular}{lll}
$\mu-\mu_0$ & $I_0 / \lambda$     & max($\rho_{\mathrm{supercrit}}$)\\\hline
$0.8$       & $4.7\times 10^{-3}$ & $3.1\times 10^{-3}$ \\
$1.2$       & $7.5\times 10^{-3}$ & $5.4\times 10^{-3}$ \\
$1.6$       & $9.7\times 10^{-3}$ & $7.1\times 10^{-3}$ \\
\end{tabular}
\end{center}
\end{table}

\begin{table}[h]
\begin{center}
\caption{Estimate of the onset of the crossover in
  the phase-transformation dynamics. Diffusion steps of a monomer
  before desorption $N(\infty)$; diffusion length of a monomer
  $l_{\mathrm{d}}=\sqrt{N(\infty)}$; average distance between clusters
  $R_0$. } \label{tab:onset}
\begin{tabular}{llllll}
$\mu-\mu_0$ & $\Delta_{\mathrm{dif}}$ & $N(\infty)$ & $l_{\mathrm{d}}$ & $R_0$ & $l_{\mathrm{d}}/R_0$ \\\hline
$0.4$ & $15$     &        &       & $14.3$ & \\
      & $10$     & $1.9$  & $1.4$ & $15.1$ & $0.09$\\
      & $5$      & $30.4$ & $5.5$ & $19.2$ & $0.29 $\\
$0.8$ & $15$     &        &       & $7.4$  & \\
      & $10$     & $2.2$  & $1.5$ & $8.9$  & $0.17$ \\ 
      & $5$      & $34.0$ & $5.8$ & $15.5$ & $0.37$\\
$1.2$ & $15$     &        &       & $5.4$  &  \\
      & $10$     & $2.4$  & $1.6$ & $7.0$  & $0.23$ \\
      & $5$      & $38.0$ & $6.2$ & $14.3$ & $0.43$ \\
$1.6$ & $15$     &        &       & $4.1$  &  \\
      & $10$     & $2.7$  & $1.6$ & $5.9$  & $0.27$ \\
      & $5$      & $42.4$ & $6.5$ & $13.4$ & $0.49$ \\
\end{tabular}
\end{center}
\end{table}

\renewcommand{\thetable}{A-\arabic{table}}
\setcounter{table}{0}  

\begin{table}[h]
\begin{center}
\caption{The particle classes for the
  adsorption/desorption step} \label{tab:part_class}
\begin{tabular}{cccc}
$s_i$ & $\sum_k{s_k}$ & $c_{\mathrm{a/d}}$ & $ \Delta \mathcal{H}$\\ \hline
$1$  & $4$  & $1$  & $2H + 8J $\\
     & $2$  & $2$  & $2H + 4J $\\
     & $0$  & $3$  & $2H $\\
     & $-2$ & $4$  & $2H -4J$\\
     & $-4$ & $5$  & $2H -8J$\\
$-1$ & $4$  & $6$  & $-2H -8J$\\
     & $2$  & $7$  & $-2H -4J$\\
     & $0$  & $8$  & $-2H $\\
     & $-2$ & $9$  & $-2H +4J$\\
     & $-4$ & $10$ & $-2H +8J$\\
\end{tabular}
\end{center}
\end{table}

\begin{table}[h]
\begin{center}
\caption{The pair classes for the
  diffusion step} \label{tab:pair_class}
\begin{tabular}{cccccc}
$s_i$ & $s_j$ & $\sum_k{s_k}$ & $\sum_l{s_l}$ & $c_{\mathrm{dif}}$ & $ \Delta \mathcal{H}$\\ \hline
$1$  & $-1$ & $-4$ & $-2$ & $1$  & $0$\\
     &      &      & $0$  & $2$  & $-4J$ \\
     &      &      & $2$  & $3$  & $-8J$ \\
     &      &      & $4$  & $4$  & $-12J$ \\
     &      & $-2$ & $-2$ & $5$  & $4J$\\
     &      &      & $0$  & $6$  & $0$ \\
     &      &      & $2$  & $7$  & $-4J$ \\
     &      &      & $4$  & $8$  & $-8J$ \\
     &      & $0$  & $-2$ & $9$  & $8J$\\
     &      &      & $0$  & $10$ & $4J$ \\
     &      &      & $2$  & $11$ & $0$ \\
     &      &      & $4$  & $12$ & $-4J$ \\
     &      & $2$  & $-2$ & $13$ & $12J$\\
     &      &      & $0$  & $14$ & $8J$ \\
     &      &      & $2$  & $15$ & $4J$ \\
     &      &      & $4$  & $16$ & $0$ \\
\end{tabular}\\
continued on next page
\end{center}
\end{table}
\addtocounter{table}{-1}
\begin{table}[h]
\begin{center}
\caption{continued} 
\begin{tabular}{cccccc}
$s_i$ & $s_j$ & $\sum_k{s_k}$ & $\sum_l{s_l}$ & $c_{\mathrm{dif}}$ & $ \Delta \mathcal{H}$\\ \hline
$-1$ & $1$  & $-2$ & $-4$ & $17$ & $0$\\
     &      &      & $-2$ & $18$ & $4J$ \\
     &      &      & $0$  & $19$ & $8J$ \\
     &      &      & $2$  & $20$ & $12J$ \\
     &      & $0$  & $-4$ & $21$ & $-4J$\\
     &      &      & $-2$ & $22$ & $0$ \\
     &      &      & $0$  & $23$ & $4J$ \\
     &      &      & $2$  & $24$ & $8J$ \\
     &      & $2$  & $-4$ & $25$ & $-8J$\\
     &      &      & $-2$ & $26$ & $-4J$ \\
     &      &      & $0$  & $27$ & $0$ \\
     &      &      & $2$  & $28$ & $4J$ \\
     &      & $4$  & $-4$ & $29$ & $-12J$\\
     &      &      & $-2$ & $30$ & $-8J$ \\
     &      &      & $0$  & $31$ & $-4J$ \\
     &      &      & $2$  & $32$ & $0$ \\
\end{tabular}
\end{center}
\end{table}

\begin{table}[h]
\begin{center}
\renewcommand{\thefootnote}{\fnsymbol{footnote}}
\setcounter{footnote}{0}
\caption{Update of the particle classes
  $c_{\mathrm{a/d}}$ after a move: $c_{\mathrm{new}} = c_{\mathrm{old}} +
  \Delta c$. For the designation of the spins and particles, see the
  text. For the meaning of the symbols for the site, see
  Fig.~\ref{fig:nn_coord}.} \label{tab:class_update_part}
\small
\begin{tabular}{llll}
move & description & $\Delta c$ &site \\
\hline
a/d & X is C & $+5s_{i,\mathrm{old}}$ & central\\
    & X is NN of C & $+s_{i,\mathrm{old}}$ & NN 0, 1, 2, 3\\
\hline
dif & X is C & $+4s_{i,\mathrm{old}}$ & central\footnotemark[1]\footnotemark[2] \\
    & X is C' & $+4s_{j,\mathrm{old}}$ & NN 1\footnotemark[1]\\
    &         &          &  NN 3\footnotemark[2]\\ 
    & NN of X is C & $+s_{i,\mathrm{old}}$ & NN 0, 2, 3\footnotemark[1]\\ 
    &              &         & NN 0, 1, 2\footnotemark[2]\\ 
    & NN of X is C' & $+s_{j,\mathrm{old}}$ & NNN 1, 2; 3NN 1\footnotemark[1]\\ 
    &               &         & NNN 1, 3; 3NN 3\footnotemark[2]\\
\multicolumn{4}{l}{\rule{0.4\textwidth}{0.5pt}}\\
\multicolumn{4}{l}{\footnotesize
\footnotemark[1]{move was diffusion to the right: C' is NN 1}}\\
\multicolumn{4}{l}{\footnotesize\footnotemark[2]{move was diffusion to site below: C' is
    NN 3}}\\
\end{tabular}
\end{center}
\end{table}

\begin{table}[h]
\begin{center}
\renewcommand{\thefootnote}{\fnsymbol{footnote}}
\setcounter{footnote}{0}
\caption{Update of the pair classes after
  an adsorption/desorption move:
  $c_{\mathrm{new}} = c_{\mathrm{old}} + \Delta c$. $m=2$: affected
  class is $c_{\mathrm{dif}}$ for  diffusion to nearest-neighbor
  position to the right (NN 1); $m=3$: affected class is
  $c_{\mathrm{dif}}$ for diffusion to 
  nearest-neighbor position below (NN 3). For the designation of the
  spins and particles, see the text. For the meaning of the symbols for
  the site, see Fig.~\ref{fig:nn_coord}.} \label{tab:class_update_pair_ad}
\small
\begin{tabular}{llll}
\multirow{2}{*}{description} & \multirow{2}{*}{$\Delta c$}
&\multicolumn{2}{c}{site} \\
& &  $m=2$ & $m=3$\\
\hline
X is C & $-32s_{i,\mathrm{old}} t_j$ & central &  central \\
X' is C & $-32s_{i,\mathrm{old}} t_i -16s_{i,\mathrm{old}}$ & NN 0 & NN 2\\ 
NN of X is C & $-4s_{i,\mathrm{old}}$ & NN 1, 2, 3 & NN 0, 1, 3\\
NN of X' is C & $-s_{i,\mathrm{old}}$ & NNN 0, 3; 3NN 0 & NNN 0, 2; 3NN 2\\ 
\end{tabular}\\
\end{center}
\end{table}

\begin{table}[h]
\begin{center}
\renewcommand{\thefootnote}{\fnsymbol{footnote}}
\setcounter{footnote}{0}
\caption{Update of the pair classes
  after a diffusion move:
  $c_{\mathrm{new}} = c_{\mathrm{old}} + \Delta c$. $m=2$: affected
  class is $c_{\mathrm{dif}}$ for  diffusion to nearest-neighbor
  position to the right (NN 1); $m=3$: affected class is
  $c_{\mathrm{dif}}$ for diffusion to 
  nearest-neighbor position below (NN 3). For the designation of the
  spins, particles and pairs, see the text. For the meaning of the symbols for
  the site, see Fig.~\ref{fig:nn_coord}.} \label{tab:class_update_pair_dif}
\small
\begin{tabular}{llll}
\multirow{2}{*}{description} &
  \multirow{2}{*}{$\Delta c$} &\multicolumn{2}{c}{site} \\
& &  $m=2$ & $m=3$\\
\hline
T is S & $+16s_{i,\mathrm{old}}$ & central\footnotemark[1] & central\footnotemark[2]\\
X is C, T $ \perp $ S & $-32s_{i,\mathrm{old}} t_j +4s_{i,\mathrm{old}}$ &
  central\footnotemark[2] & central\footnotemark[1] \\
X' is C & $-32s_{i,\mathrm{old}} t_i-15s_{i,\mathrm{old}}$ & NN 0\footnotemark[1]\footnotemark[2]
  & NN 2\footnotemark[1]\footnotemark[2]\\
X is C' & $-32s_{j,\mathrm{old}} t_j + 4 s_j$ & NN 1\footnotemark[1]& NN 1\footnotemark[1]\\  
&  & NN 3\footnotemark[2] & NN 3\footnotemark[2] \\ 
X' is C' & $-32s_{j,\mathrm{old}} t_i -15s_{j,\mathrm{old}}$ &NNN 3\footnotemark[2]
  & NNN 2\footnotemark[1]\\  
T $||$ S & $-3s_{i,\mathrm{old}}$ & NN 2, 3\footnotemark[1] & NN 0, 1\footnotemark[2] \\
NN of X is C & $-4s_{i,\mathrm{old}}$ & NN 1, 2\footnotemark[2] & NN 0, 3\footnotemark[1]\\
NN of X' is C & $-s_{i,\mathrm{old}}$ & NNN 0, 3; 3NN 0\footnotemark[1] & NNN 1; 3NN
  1\footnotemark[1]\\
&  & NNN 0; 3NN 0\footnotemark[2] &  NNN 1, 3; 3NN 3\footnotemark[2]\\
NN of X is C' & $-s_{i,\mathrm{old}}$ & NNN 1, 2; 3NN 1\footnotemark[1] & NNN 0; 3NN
  2\footnotemark[1] \\
&  & NNN 1; 3NN 3\footnotemark[2] &  NNN 0, 2; 3NN 2\footnotemark[2]\\
NN of X' is C' & $-s_{i,\mathrm{old}}$ & 4NN 0, 3\footnotemark[2] & 4NN 1, 2\footnotemark[1]\\ 

\multicolumn{4}{l}{\rule{0.4\textwidth}{0.5pt}}\\
\multicolumn{4}{l}{\footnotesize
\footnotemark[1]{move was diffusion to the right: C' is NN 1}}\\
\multicolumn{4}{l}{\footnotesize\footnotemark[2]{move was diffusion to site below: C' is
    NN 3}}\\
\end{tabular}\\
\end{center}
\end{table}


\clearpage
\newpage
\renewcommand{\thefigure}{\arabic{figure}}
\setcounter{figure}{0}

\begin{figure}[h]
\includegraphics[width=\textwidth, clip]{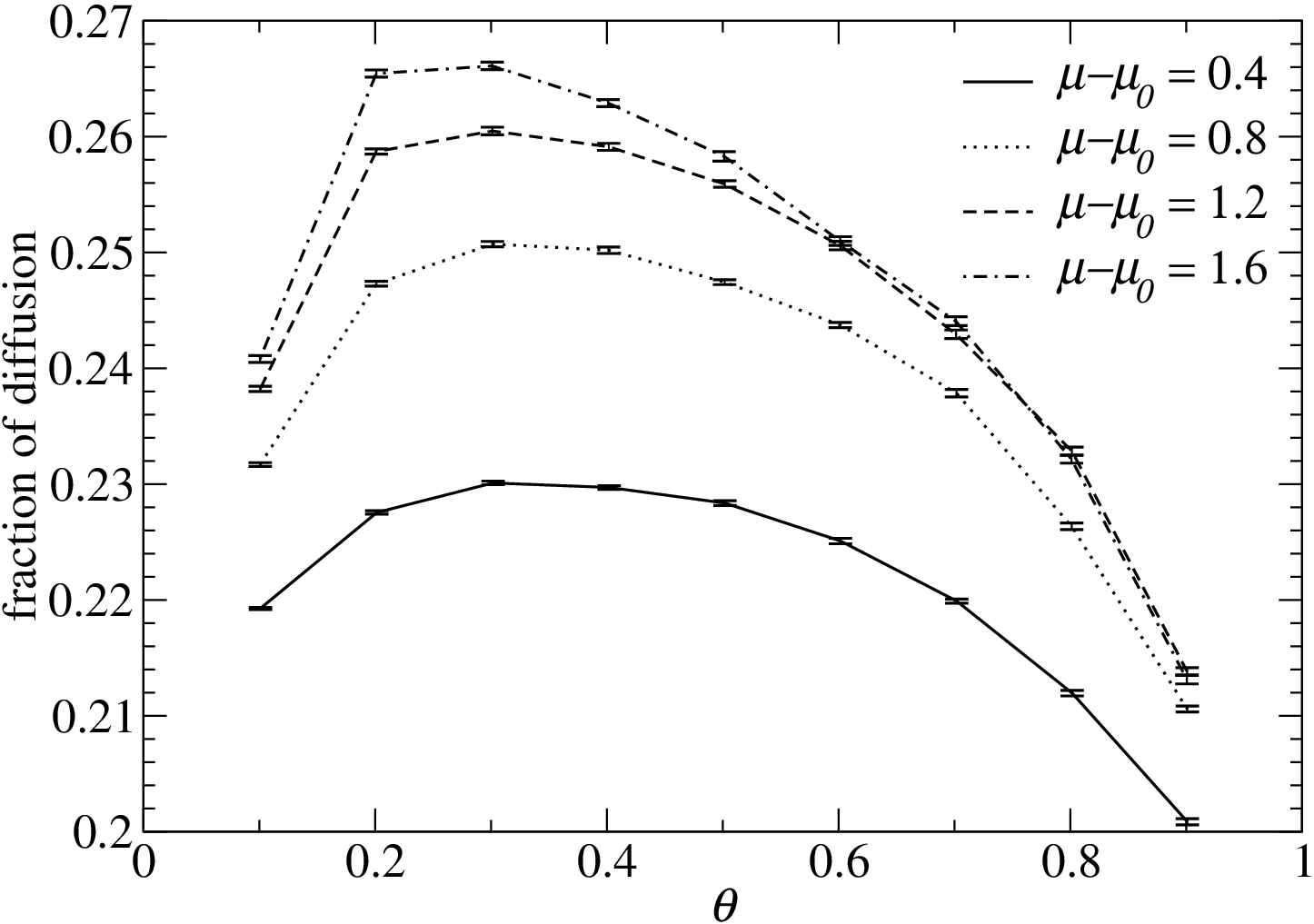}
\caption{}
\end{figure}

\begin{figure}[h]
\includegraphics[width=\textwidth, clip]{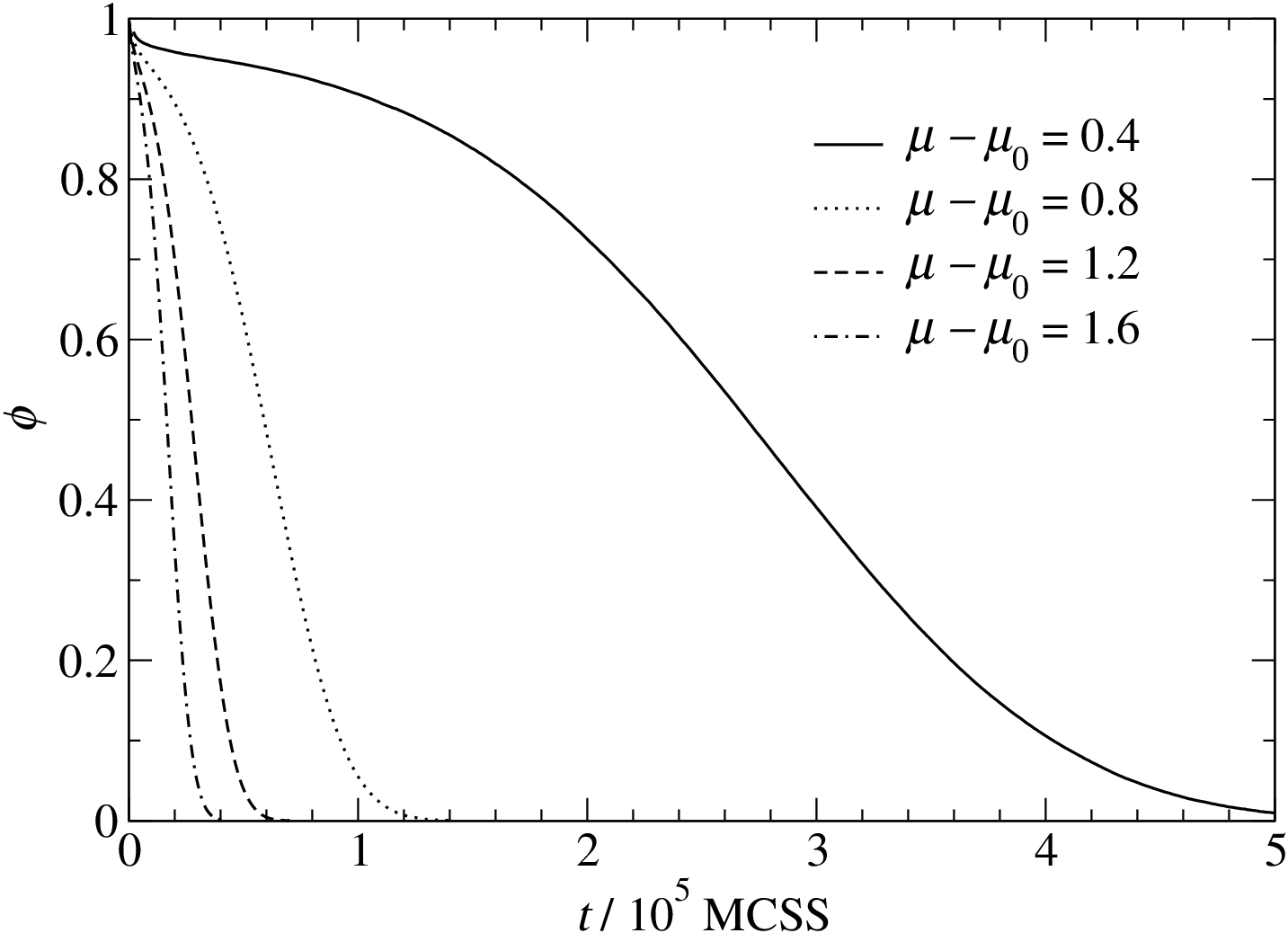}
\caption{}
\end{figure}

\begin{figure}[h]
\includegraphics[width=\textwidth, clip]{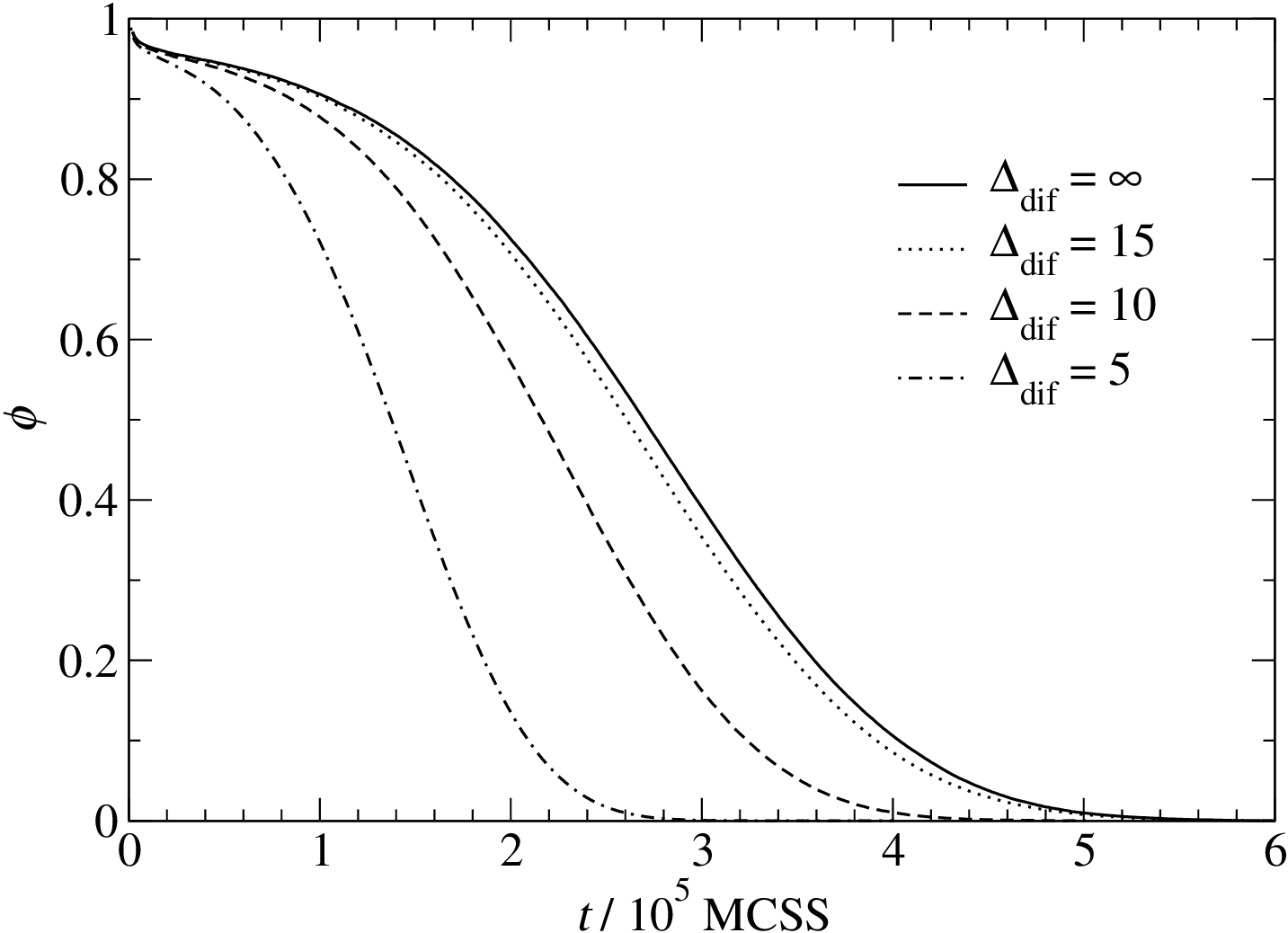}
\caption{}
\end{figure}

\begin{figure}[h]
\includegraphics[width=\textwidth, clip]{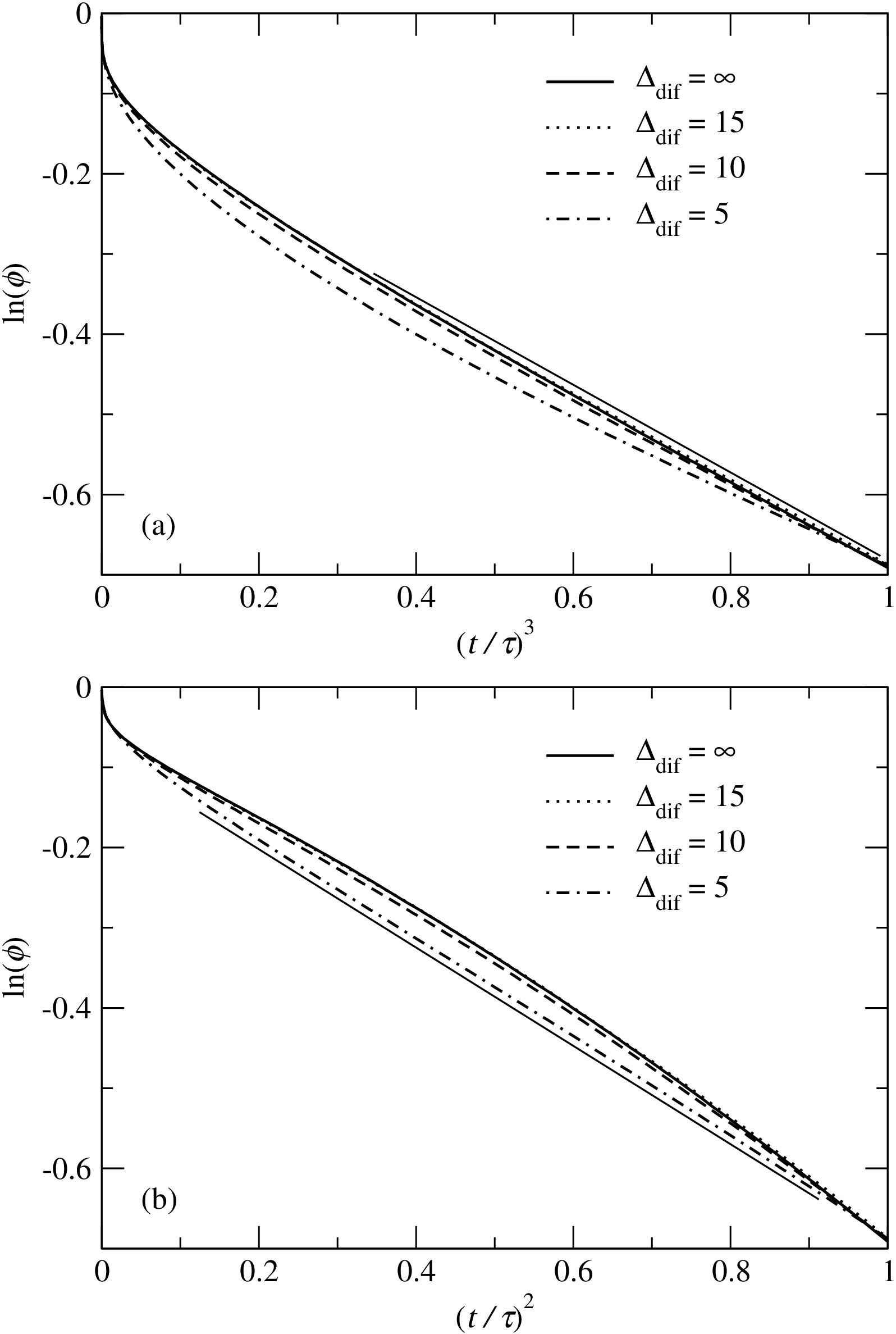}
\caption{}
\end{figure}

\begin{figure}[h]
\includegraphics[width=\textwidth, clip]{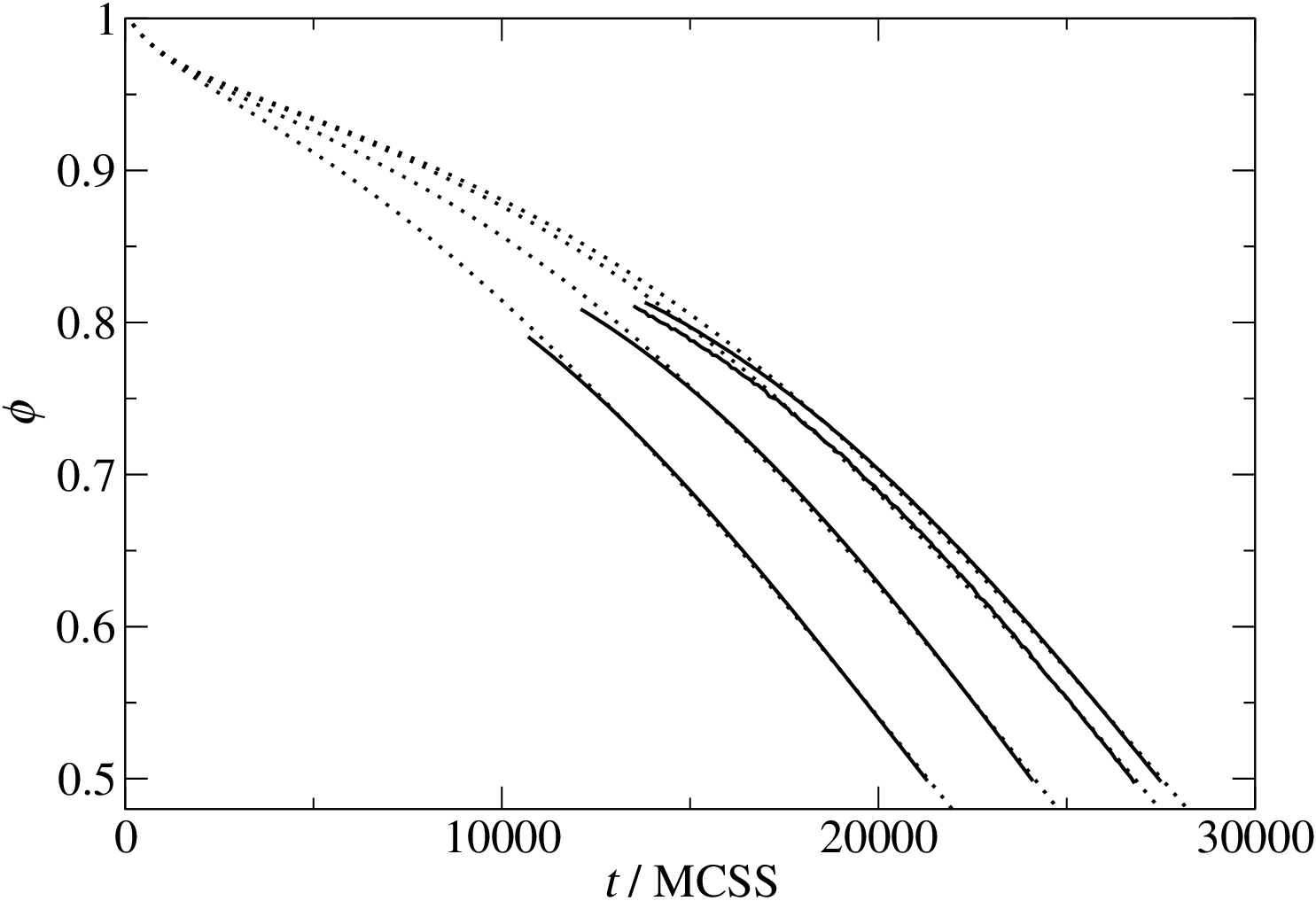}
\caption{}
\end{figure}

\begin{figure}[h]
\includegraphics[width=\textwidth, clip]{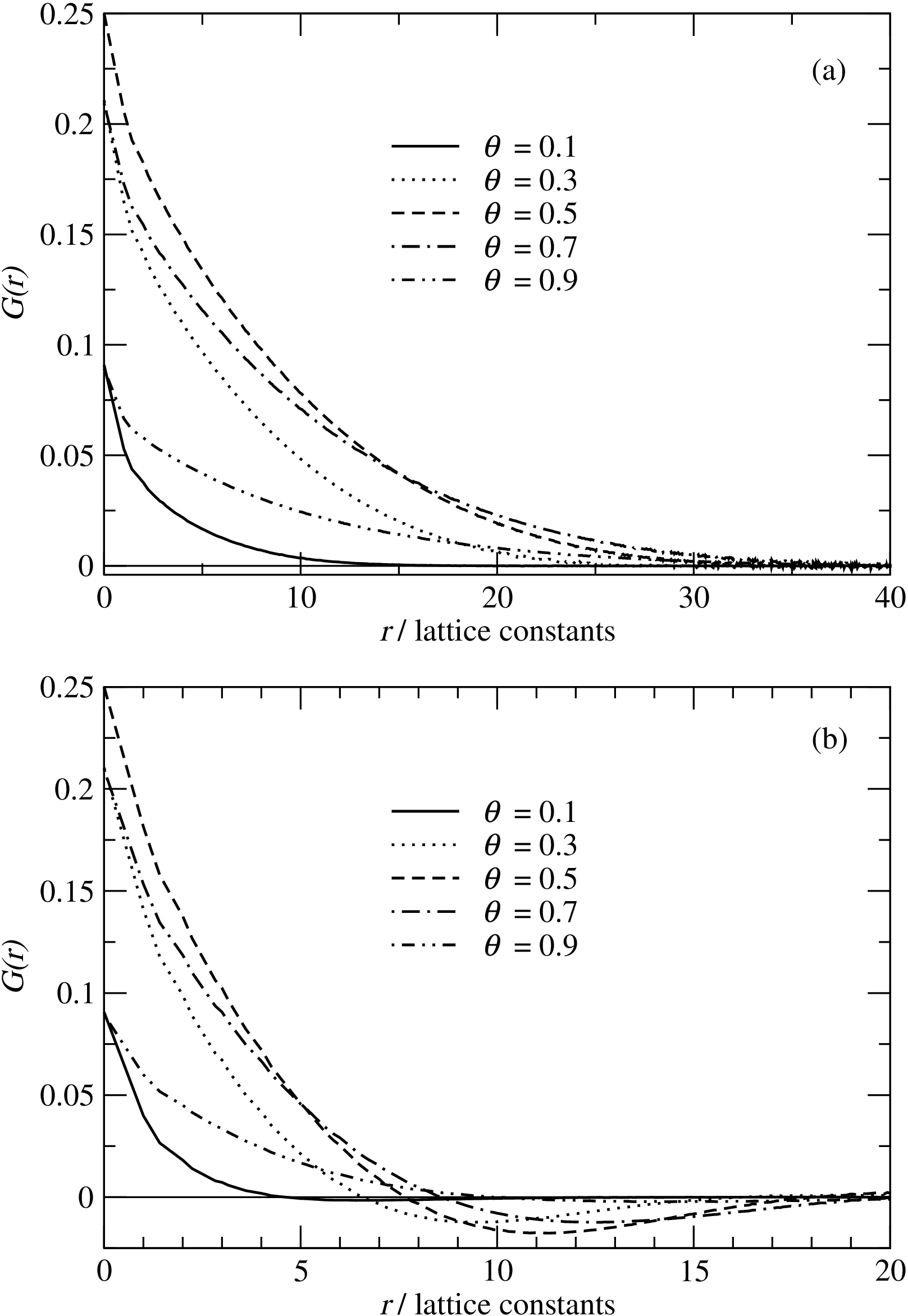}
\caption{}
\end{figure}

\begin{figure}[h]
\includegraphics[width=\textwidth, clip]{Fig07.eps}
\caption{}
\end{figure}

\begin{figure}[h]
\includegraphics[width=\textwidth, clip]{Fig08.eps}
\caption{}
\end{figure}

\begin{figure}[h]
\includegraphics[width=\textwidth, clip]{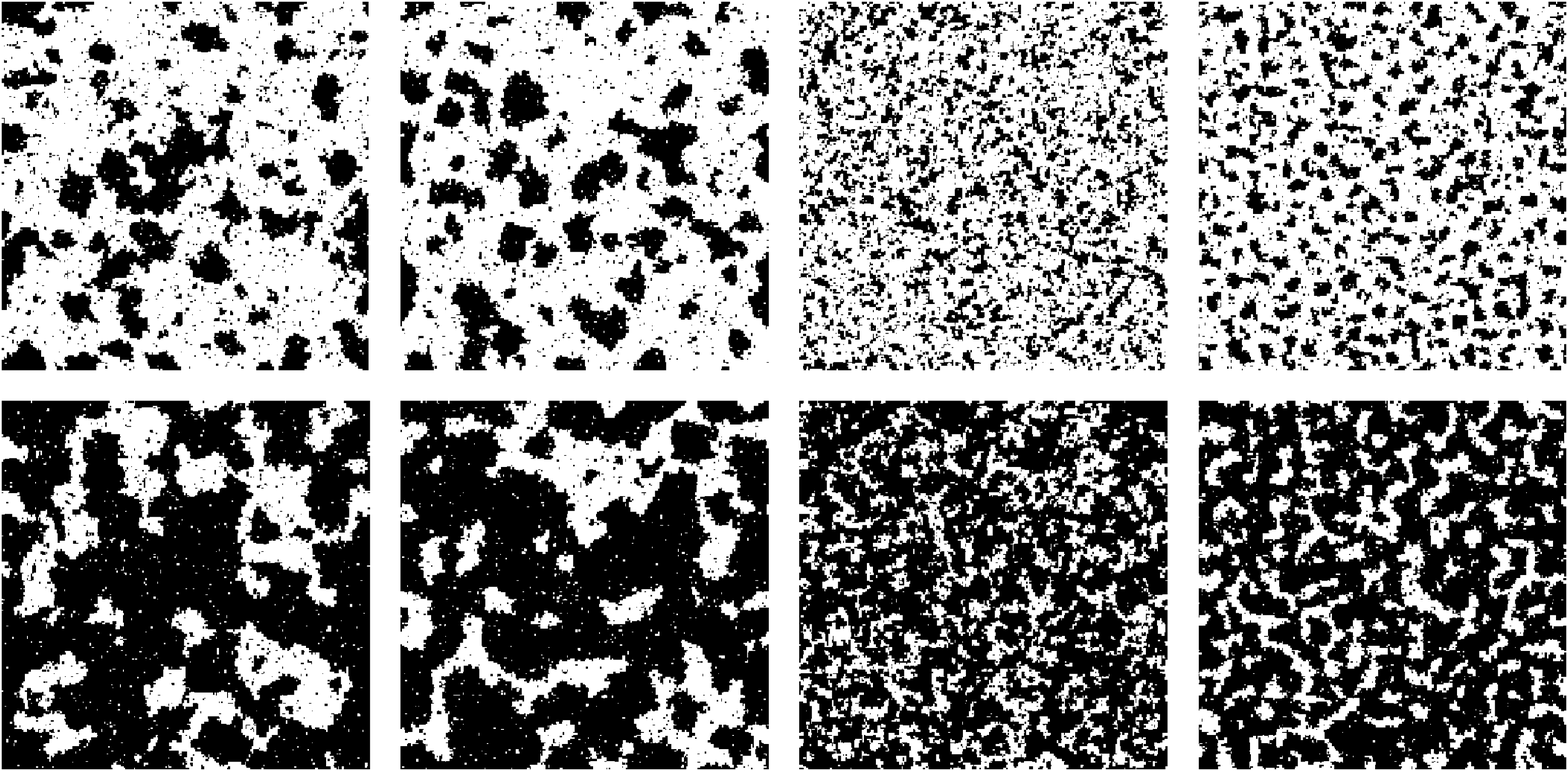}
\caption{}
\end{figure}

\begin{figure}[h]
\includegraphics[width=\textwidth, clip]{Fig10.eps}
\caption{}
\end{figure}

\begin{figure}[h]
\includegraphics[width=\textwidth, clip]{Fig11.eps}
\caption{}
\end{figure}

\begin{figure}[h]
\includegraphics[width=\textwidth, clip]{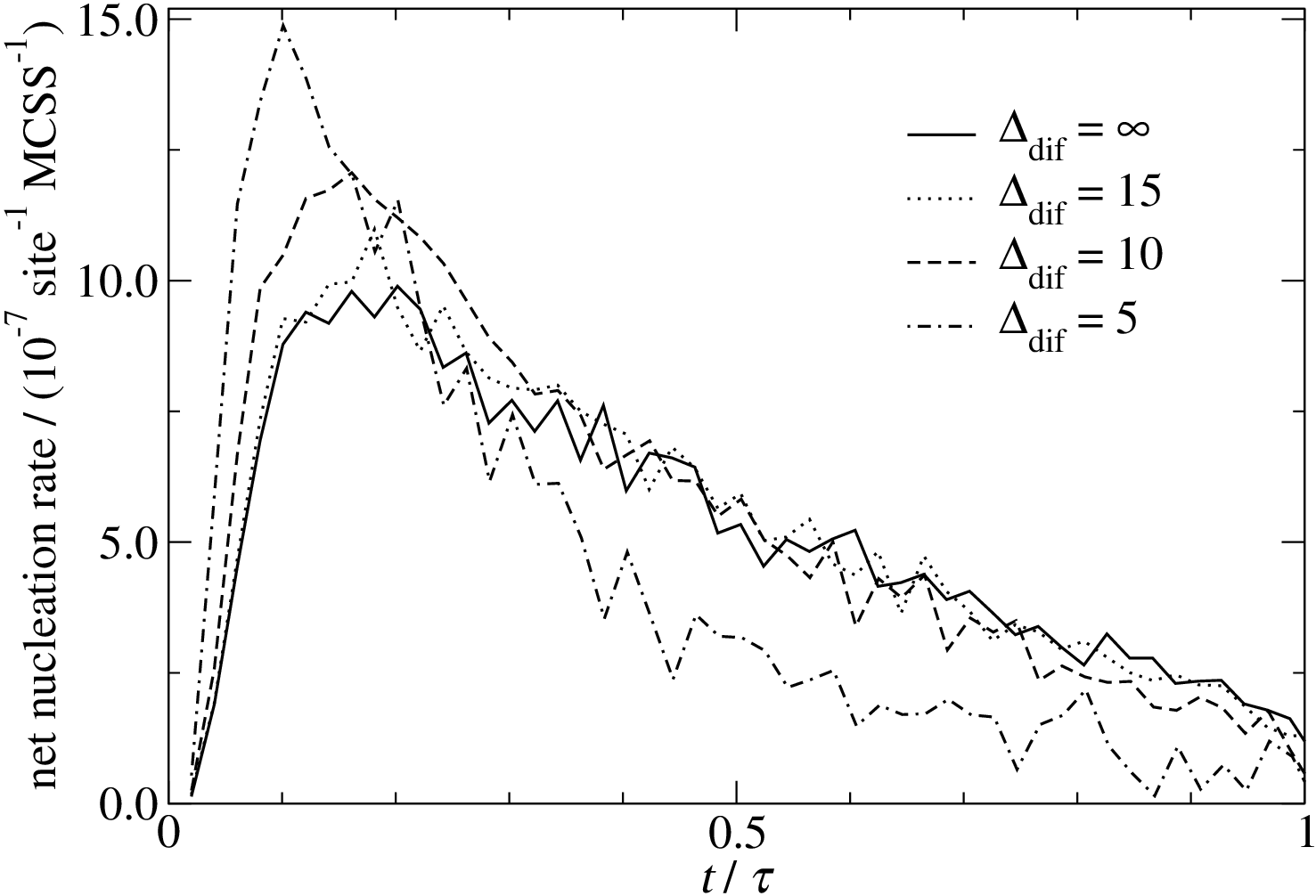}
\caption{} 
\end{figure}

\begin{figure}[h]
\includegraphics[width=\textwidth, clip]{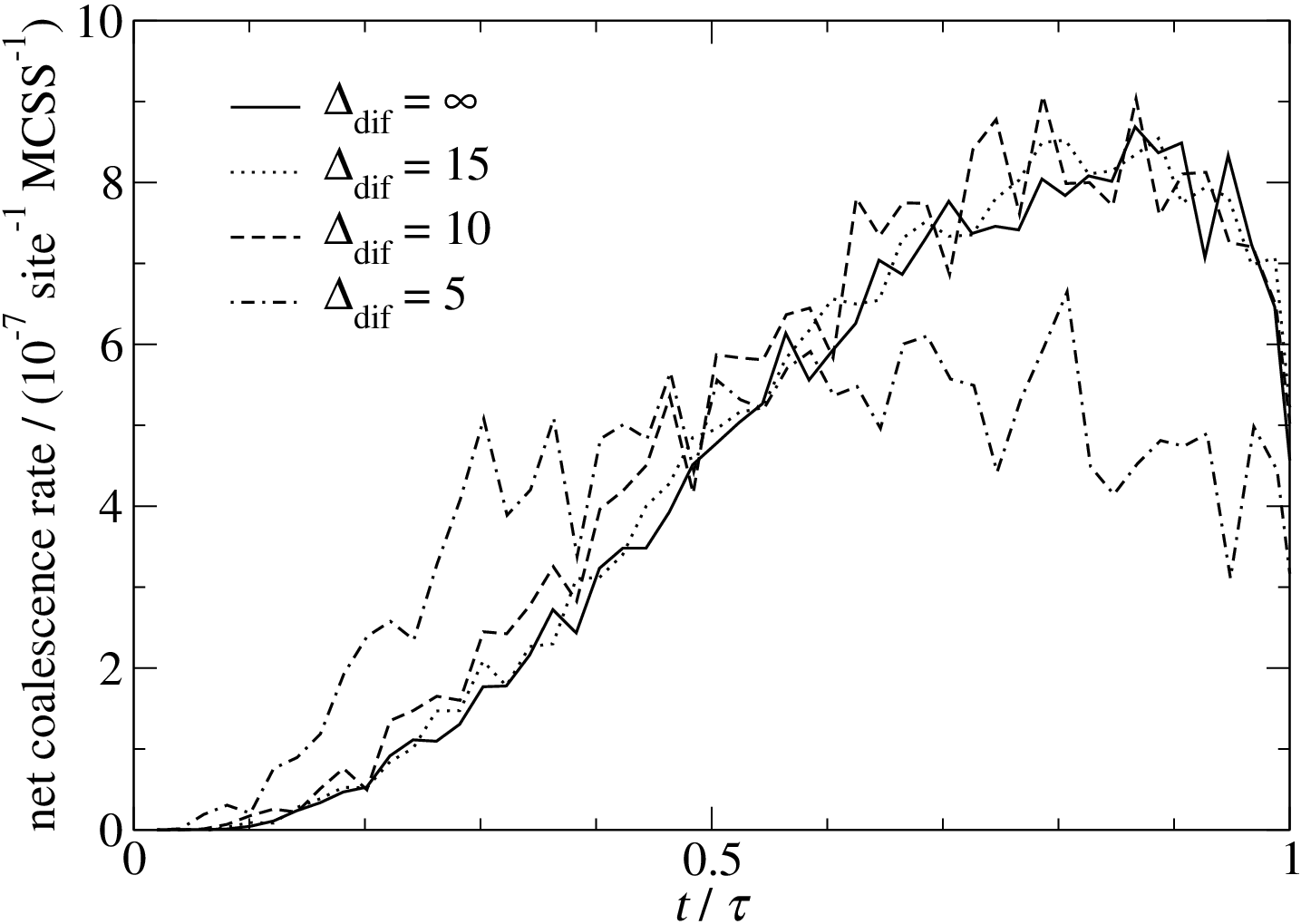}
\caption{}
\end{figure}

\begin{figure}[h]
\includegraphics[width=\textwidth, clip]{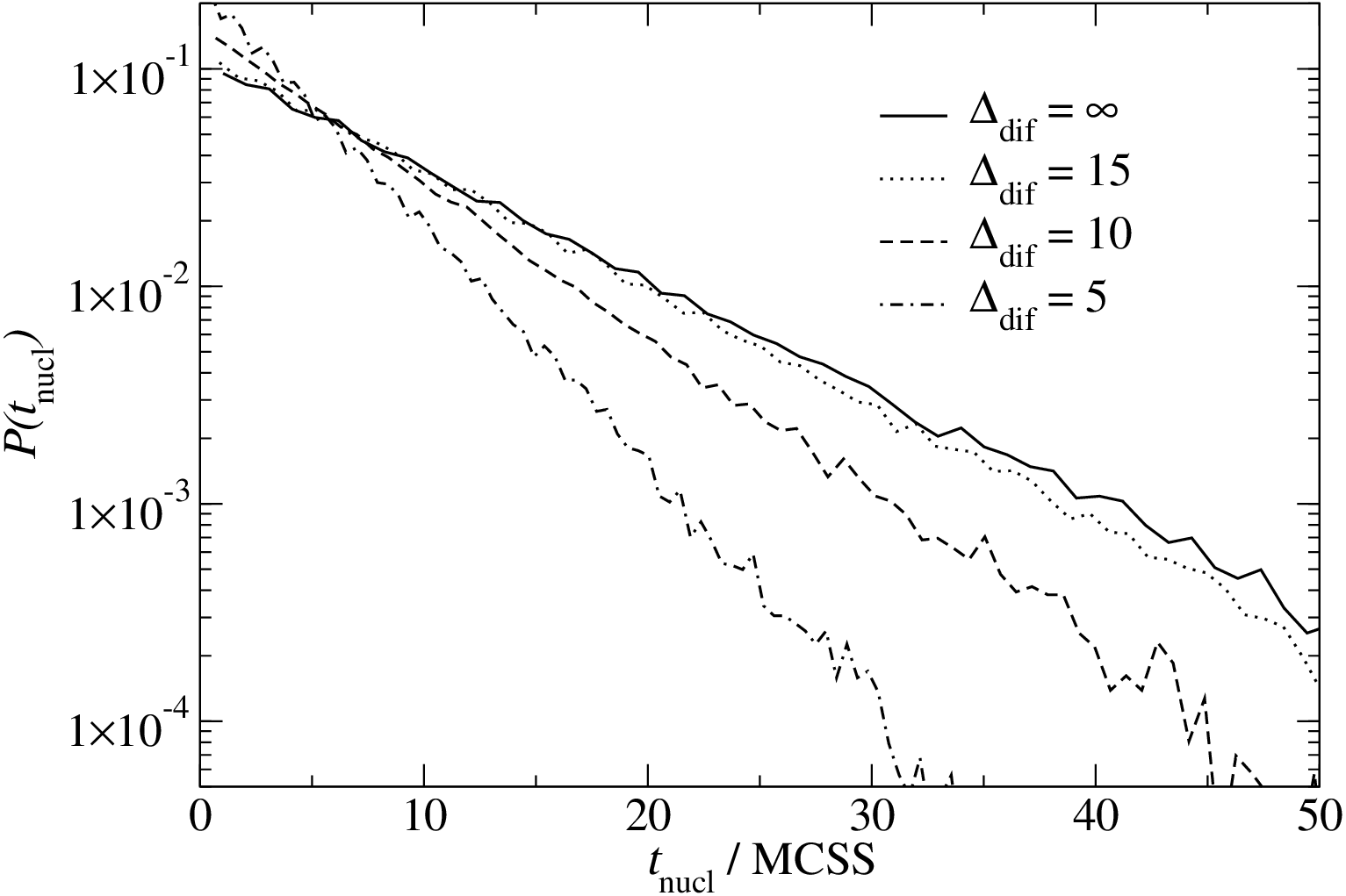}
\caption{}
\end{figure}

\begin{figure}[h]
\includegraphics[width=\textwidth, clip]{Fig15.eps}
\caption{}  
\end{figure}

\clearpage
\renewcommand{\thefigure}{A-\arabic{figure}}
\setcounter{figure}{0}  

\begin{figure}[h]
\includegraphics[width=\textwidth,clip]{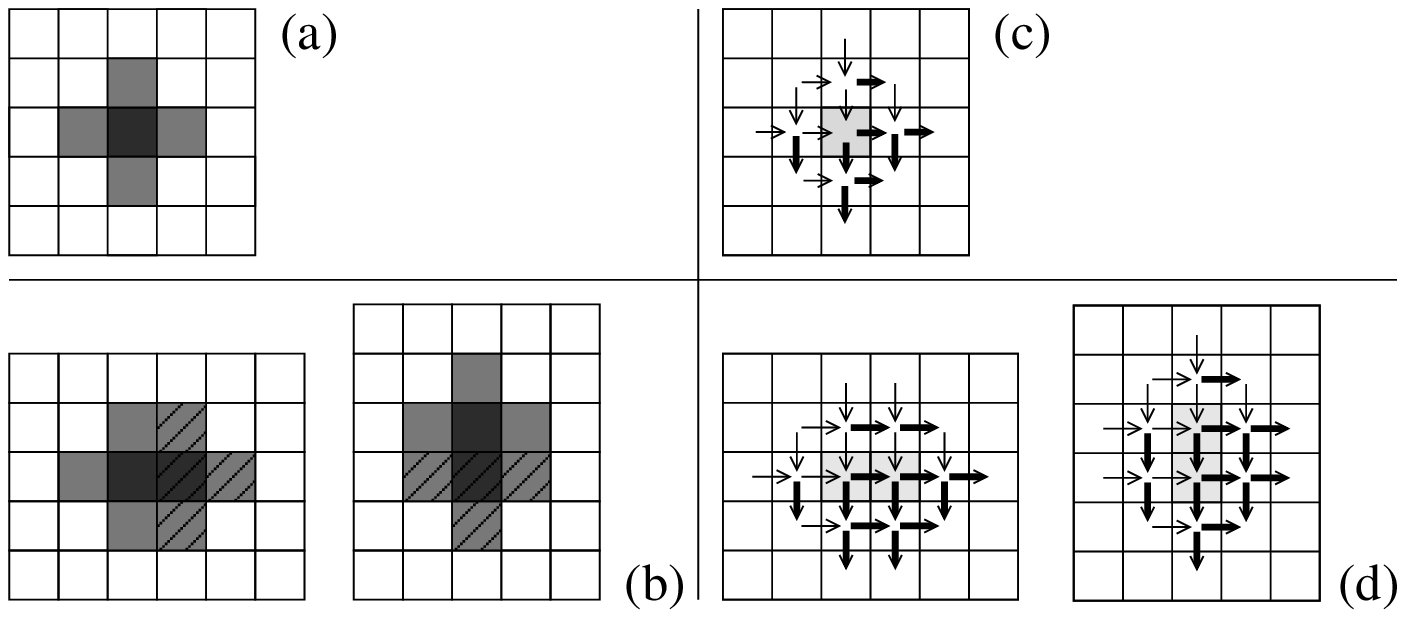}
\caption{(greyscale)}
\end{figure}
\addtocounter{figure}{-1}
\begin{figure}[h]
\includegraphics[width=\textwidth,clip]{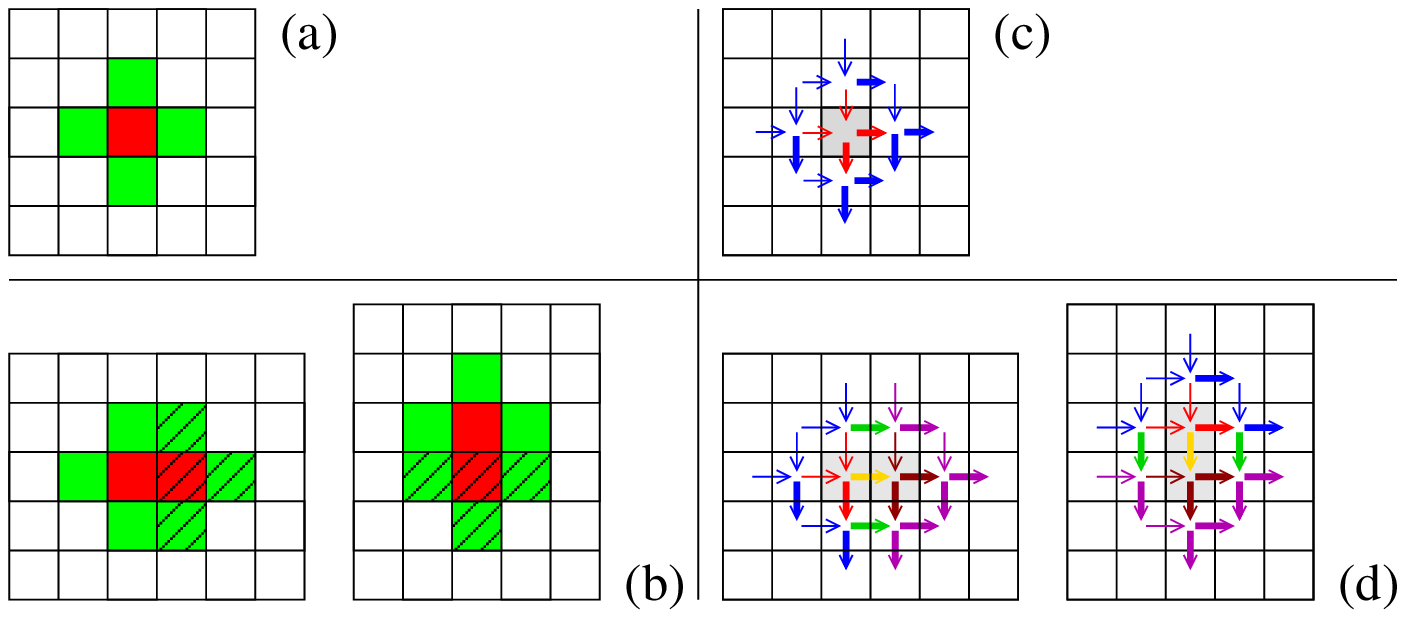}
\caption{(color)}
\end{figure}

\begin{figure}[h]
\includegraphics[width=\textwidth,clip]{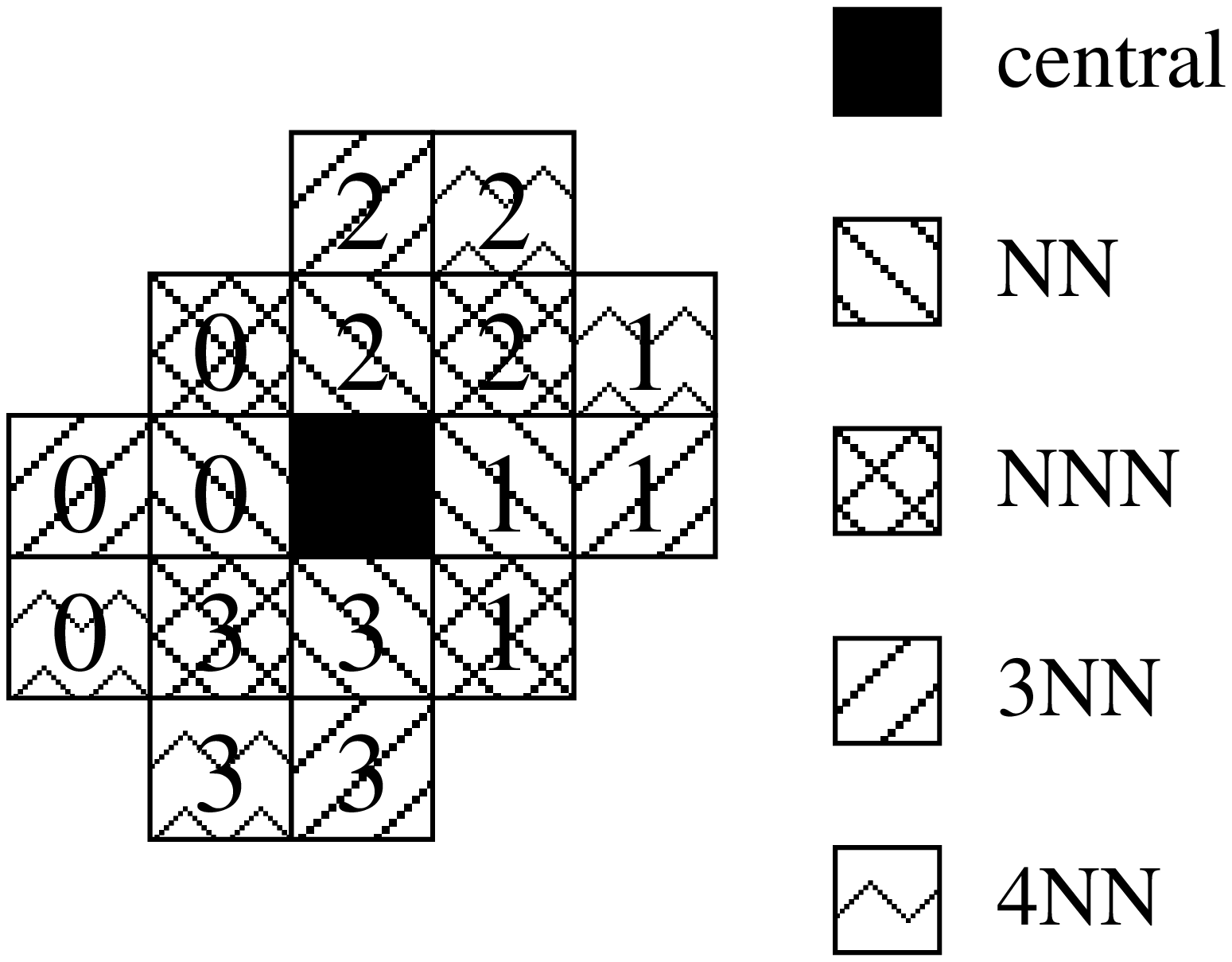}
\caption{(greyscale)}
\end{figure}
\addtocounter{figure}{-1}
\begin{figure}[h]
\includegraphics[width=\textwidth,clip]{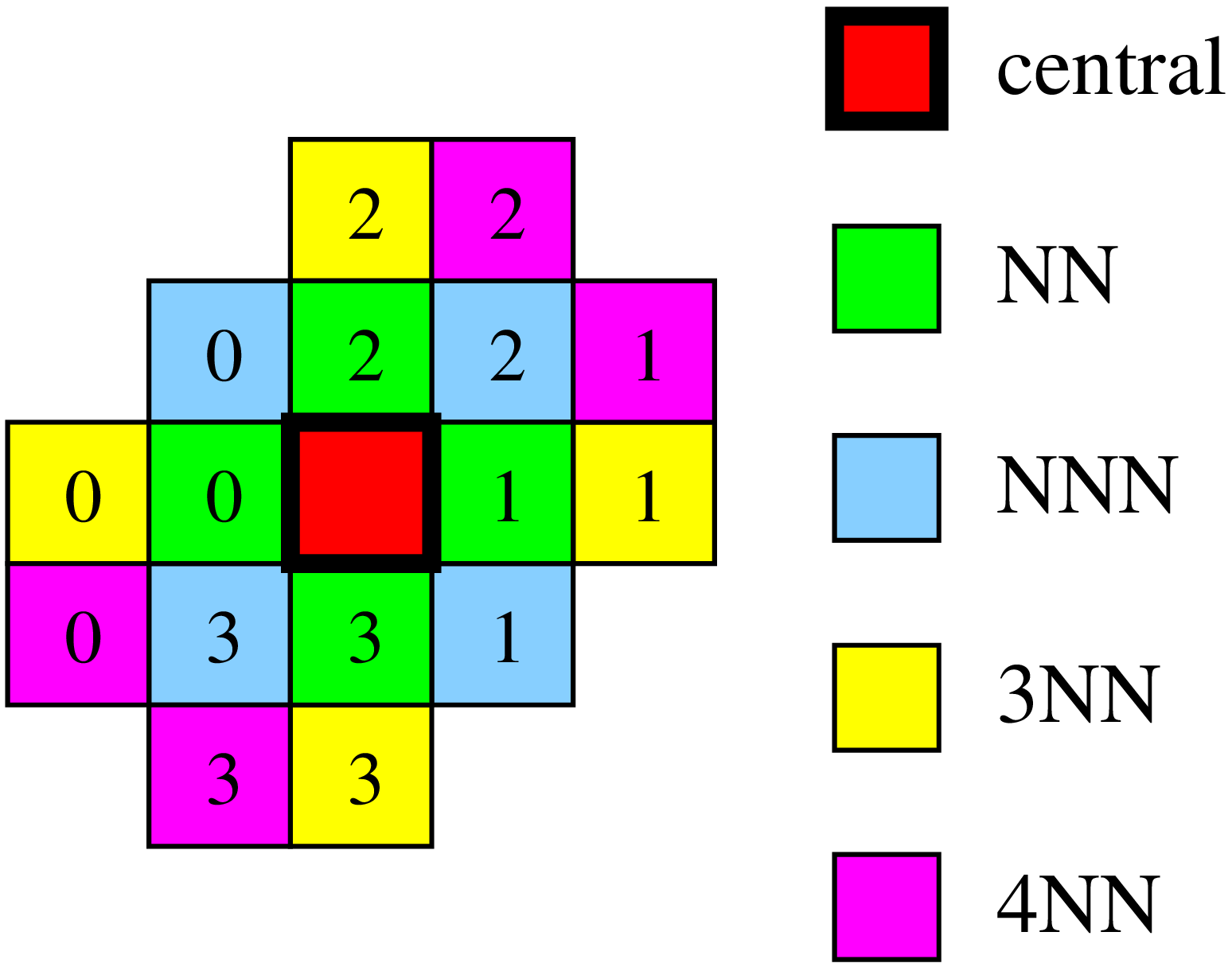}
\caption{(color)}
\end{figure}

\end{document}